\documentclass[8.5pt,twoside,twocolumn]{article}
\usepackage[super,sort&compress,comma]{natbib} 
\usepackage{mhchem}
\usepackage{times,mathptmx}
\DeclareMathAlphabet{\mathcal}{OMS}{cmsy}{m}{n}
\usepackage{sectsty}
\usepackage{balance} 
\usepackage{epstopdf}
\usepackage{graphicx} 
\usepackage{lastpage}
\usepackage[format=plain,justification=raggedright,singlelinecheck=false,font=small,labelfont=bf,labelsep=space]{caption} 
\usepackage{fancyhdr}
\begin{document}

\thispagestyle{plain}
\fancypagestyle{plain}{
\fancyhead[L]{\includegraphics[height=8pt]{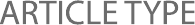}}
\fancyhead[C]{\hspace{-1cm}\includegraphics[height=20pt]{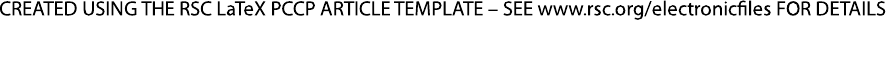}}
\fancyhead[R]{\includegraphics[height=10pt]{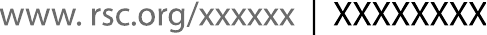}\vspace{-0.2cm}}
\renewcommand{\headrulewidth}{1pt}}
\renewcommand{\thefootnote}{\fnsymbol{footnote}}
\renewcommand\footnoterule{\vspace*{1pt}%
\hrule width 3.4in height 0.4pt \vspace*{5pt}} 
\setcounter{secnumdepth}{5}

\makeatletter 
\def\subsubsection{\@startsection{subsubsection}{3}{10pt}{-1.25ex plus -1ex minus -.1ex}{0ex plus 0ex}{\normalsize\bf}} 
\def\paragraph{\@startsection{paragraph}{4}{10pt}{-1.25ex plus -1ex minus -.1ex}{0ex plus 0ex}{\normalsize\textit}} 
\renewcommand\@biblabel[1]{#1}            
\renewcommand\@makefntext[1]%
{\noindent\makebox[0pt][r]{\@thefnmark\,}#1}
\makeatother 
\renewcommand{\figurename}{\small{Fig.}~}
\sectionfont{\large}
\subsectionfont{\normalsize} 

\fancyfoot{}
\fancyfoot[LO,RE]{\vspace{-7pt}\includegraphics[height=9pt]{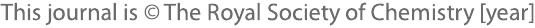}}
\fancyfoot[CO]{\vspace{-7.2pt}\hspace{12.2cm}\includegraphics{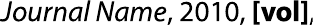}}
\fancyfoot[CE]{\vspace{-7.5pt}\hspace{-13.5cm}\includegraphics{RF}}
\fancyfoot[RO]{\footnotesize{\sffamily{1--\pageref{LastPage} ~\textbar  \hspace{2pt}\thepage}}}
\fancyfoot[LE]{\footnotesize{\sffamily{\thepage~\textbar\hspace{3.45cm} 1--\pageref{LastPage}}}}
\fancyhead{}
\renewcommand{\headrulewidth}{1pt} 
\renewcommand{\footrulewidth}{1pt}
\setlength{\arrayrulewidth}{1pt}
\setlength{\columnsep}{6.5mm}
\setlength\bibsep{1pt}

\twocolumn[
  \begin{@twocolumnfalse}
\noindent\LARGE{\textbf{Native point defects in CuIn$_{1-x}$Ga$_x$Se$_{2}$: hybrid density functional calculations predict origin of p- and n-type conductivity}}
\vspace{0.6cm}

\noindent\large{\textbf{J.~Bekaert,$^{\ast}$ R.~Saniz, B.~Partoens and
D.~Lamoen }}\vspace{0.5cm}

\noindent\textit{\small{\textbf{Received Xth XXXXXXXXXX 20XX, Accepted Xth XXXXXXXXX 20XX\newline
First published on the web Xth XXXXXXXXXX 200X}}}

\noindent \textbf{\small{DOI: 10.1039/b000000x}}
\vspace{0.6cm}

\noindent \normalsize{We have performed a first-principles study of the p- and n-type conductivity in CuIn$_{1-x}$Ga$_x$Se$_{2}$ due to native point defects, based on the HSE06 hybrid functional. Band alignment shows that the band gap becomes larger with $x$ due to the increasing conduction band minimum, rendering it hard to establish n-type conductivity in CuGaSe$_{2}$. From the defect formation energies, we find that In/Ga$_{\mathrm{Cu}}$ is a shallow donor, while V$_{\mathrm{Cu}}$, V$_{\mathrm{In}/\mathrm{Ga}}$ and Cu$_{\mathrm{In}/\mathrm{Ga}}$ act as shallow acceptors. Using total charge neutrality of ionized defects and intrinsic charge carriers to determine the Fermi level, we show that under In-rich growth conditions In$_{\mathrm{Cu}}$ causes strongly n-type conductivity in CuInSe$_{2}$. Under In-poor growth conditions the conductivity type in CuInSe$_{2}$ alters to p-type and compensation of the acceptors by In$_{\mathrm{Cu}}$ reduces, as observed in photoluminescence experiments. In CuGaSe$_{2}$, the native acceptors pin the Fermi level far away from the conduction band minimum, thus inhibiting n-type conductivity. On the other hand, CuGaSe$_{2}$ shows strong p-type conductivity under a wide range of Ga-poor growth conditions. Maximal p-type conductivity in CuIn$_{1-x}$Ga$_x$Se$_{2}$ is reached under In/Ga-poor growth conditions, in agreement with charge concentration measurements on samples with In/Ga-poor stoichiometry, and is primarily due to the dominant acceptor Cu$_{\mathrm{In}/\mathrm{Ga}}$.
}
\vspace{0.5cm}
 \end{@twocolumnfalse}
  ]



\footnotetext{\textit{CMT-group and EMAT, Department of Physics, University of Antwerp,\\ Groenenborgerlaan 171, B-2020 Antwerp, Belgium.\\ E-mail: jonas.bekaert@uantwerpen.be}}


\section{Introduction}
CuIn$_{1-x}$Ga$_x$Se$_{2}$ (CIGS) is a I-III-VI$_2$ semiconductor compound, where $0 \leq x \leq 1$ denotes the Ga-to-In ratio. Thus, it can be considered a semiconductor alloy of CuInSe$_2$ (CIS) and CuGaSe$_2$ (CGS). Both adopt the body-centered tetragonal chalcopyrite structure, characterized by space group I$\bar{4}$2d, in their ground state. CIGS is of particular interest as the absorber material in thin-film photovoltaic cells, as it has a very high optical absorption coefficient \cite{Kodigala}. Moreover, photovoltaic cells based on polycrystalline CIGS hold record conversion efficiencies in the category of thin-film cells, currently already exceeding 20\%, both on glass substrates and on flexible substrates \cite{Jackson,Chirila}. Photoluminescence spectra of CIS and CGS, \textit{e.g.}~in Refs.~\citenum{Bauknecht,Siebentritt2004,Siebentritt13}, are used to study the mechanisms behind the conductivity in these materials. The photoluminescence spectra show three separate donor-acceptor transitions in Cu-rich samples, but a single, broadened peak in Cu-poor samples. In the latter case, the peak is asymmetric and subject to a red-shift as the sample is made more Cu-poor. From these observations, the conclusion can be drawn that potential fluctuations are present in Cu-poor samples, caused by a strong compensation of acceptors and donors. On the other hand, there is much less compensation in Cu-rich samples. In these samples, the intensity of the peak due to the most shallow acceptor decreases with an increasing amount of Cu, whereas that of the peak due to the second acceptor increases. The third peak, due to the least shallow acceptor, keeps more or less a constant intensity. In addition, the hole concentration is found to increase with the stoichiometry $\left[\mathrm{Cu}\right]/\left[\mathrm{In}\right]$ and $\left[\mathrm{Cu}\right]/\left[\mathrm{Ga}\right]$ in CIS and CGS respectively \cite{Siebentritt13,Siebentritt01}. The measured hole concentration increases likewise with the Ga-to-In ratio, while there is also a difference in freeze out behavior \cite{Schroeder,Schoen2000}. Ref.~\citenum{Schoen2000} shows a charge carrier freeze out of less than one order of magnitude upon cooling from 500 to 50 K in CGS, while in CIS it is around two orders of magnitude. Although these experimental studies clearly give important information about the effect of the experimental growth conditions, they do not produce atomic-scale information of the point defects that give rise to the observed conductivity. For this purpose, first-principles calculations, based on the density functional theory (DFT) formalism, prove useful. From first-principles, C.~Persson \textit{et al.}~\cite{Persson2005} have found a strong tendency (in terms of the formation energy of the defect) to form the In$_{\mathrm{Cu}}$ antisite defect in CIS and the Ga$_{\mathrm{Cu}}$ antisite defect in CGS, under Se-poor growth conditions. On the other hand, in publications by S.-H.~Wei \textit{et al.}~\cite{Wei1998,Wei2013}, under similar growth conditions, the In$_{\mathrm{Cu}}$ and Ga$_{\mathrm{Cu}}$ antisite defects as point defects (\textit{i.e.}~not present in clusters with other defects) are predicted to be much less prevalent. In the latter case, the question arises how to explain the n-type conducting CIS observed experimentally \cite{Kodigala}. The aforementioned first-principles studies make use of the local density approximation (LDA) to DFT, an approach with significant limitations for studying semiconductor materials. Namely, the band gaps of semiconductors (and insulators as well) are systematically underestimated \cite{Sham1983}. As described in Ref.~\citenum{Persson2005}, calculations of defects within LDA require several a posteriori corrections. To overcome the band gap problem in a more natural way, we employ the hybrid functional method (\textit{i.e.}~combining DFT with exact exchange from the Hartree-Fock method, at short distances) in our calculations, more specifically the HSE06 functional \cite{Heyd}. For similar reasons, the HSE06 functional has previously been employed in first-principles studies on CIGS by L.~E.~Oikkonen \textit{et al.}~\cite{Oikkonen2012}, J.~Pohl \textit{et al.}~\cite{Pohl2013} and B.~Huang \textit{et al.}~\cite{Huang2014}. We start our study elaborating on how the band gap changes from around 1.0 eV for CIS to 1.7 eV for CGS (as measured in optical measurements, \textit{e.g.}~in Ref.~\citenum{Tinoco}), by considering the band alignment of CIS, CGS and intermediate compounds. The band alignment is directly related to the study of point defects, as the formation energies of charged defects depend on the Fermi level in the band gap (due to the exchange of electrons). If a rise of the conduction band minimum (CBM) is the main contribution to the larger band gap - as we will demonstrate is the case upon increasing $x$ in CIGS - then it becomes harder to establish n-type doping. To determine which point defects are responsible for the conductivity, we have calculated the formation energies of several point defects, both vacancies and antisite defects. These calculations predict both shallow donors and acceptors with low formation energies. Therefore, the conductivity type and charge concentration is highly sensitive to the chemical growth conditions. We use the formation energies to predict the conductivity type and to give an estimate of the free charge carrier concentration under different conditions. In order to do this, we solve the self-consistent dependence, through charge neutrality, of the Fermi level and the defect concentrations that follow from the formation energies. This approach is rarely followed in other first-principles studies of defects, but in the case of CIGS it has also been attempted by C.~Persson \textit{et al.}~based on formation energies obtained within LDA \cite{Persson2005}. Also, J.~Pohl \textit{et al.}~give a qualitative estimation of the Fermi level - namely the level where the formation energies of the dominant acceptor and donor are equal, but did not calculate the corresponding free charge carrier concentration \cite{Pohl2013}. The determination of the Fermi level allows us to resolve the questions raised by the experiments. Namely, which native point defects are at play in the different cases and how can the similarities and differences between CIS and CGS be explained.  

\section{Band alignment as a function of Ga-to-In ratio}

We aim to set the band structures of CIS, CGS and intermediate compounds on a common energy scale. This band alignment does not directly follow from the DFT calculations, as these do not use an absolute energy level. Therefore, several techniques for band alignments have been developed, among which we have applied and compared two different ones, \textit{viz.}~(i) an alignment via slabs and (ii) an alignment using the branch-point energy. The former method, also described in Ref.~\citenum{Moses}, relies on the construction of a slab that is sufficiently thick, so the potential within the slab can be linked to the potential in the bulk material. We have found that for CIGS the slab consisting of 9 atomic layers, terminated on both sides by the (001) planes consisting of Cu and In/Ga and surrounded by an amount of vacuum of twice the thickness of the slab, meets this requirement. The main advantage of the alignment via slabs is that it contains an absolute reference: the potential in vacuum, hence also giving the electron affinities. The other method, using the concept of the branch-point energy ($E_{BP}$), was proposed by A.~Schleife \textit{et al.}~\cite{Schleife} and solely relies on the band structures of the bulk materials. The $E_{BP}$ is calculated as an average of the electronic eigenvalues over the Brillouin zone, defined in Ref.~\citenum{Schleife}.

\subsection{Computational details}

As we have mentioned in the Introduction, our calculations make use of the HSE hybrid functional approach, more specifically the HSE06 functional, as implemented in the VASP code \cite{Kresse,Paier}. Electron-ion interactions are treated using projector augmented wave (PAW) potentials, taking into account Cu-3d$^{10}$4s$^1$, Ga-3d$^{10}$4s$^2$4p$^1$, In-4d$^{10}$5s$^2$5p$^1$ and Se-4s$^2$4p$^4$ as valence electrons. The energy cutoff for the plane-wave basis is set to 500 eV. We model the alloys of CIS and CGS by means of the $1 \times 1\times 2$ supercell spanned by the vectors $\textbf{a}_1=\left(a,0,0\right)$, $\textbf{a}_2=\left(0,a,0\right)$ and $\textbf{a}_3=\left(a,a,c\right)$, where $a$ and $c$ are the lattice parameters, so $x=0,~0.25,~0.5,~0.75,~1$ can be studied. For integration over the Brillouin zone in the bulk structures a $4 \times 4\times 4$ $\Gamma$-centered Monkhorst-Pack \textbf{k}-point grid is used and scaled appropriately for the slabs to $4 \times 4 \times 1$. The integration is facilitated by Gaussian smearing with a width of $\sigma=0.05$ eV. The standard HSE06 functional (with an amount $\alpha=0.25$ of Hartree-Fock exact exchange at the short range) produces band gaps of 0.85 eV and 1.37 eV, for CIS and CGS respectively. This suggests that the exchange interaction is still overscreened with $\alpha=0.25$, resulting in smaller gaps compared to experiment. The agreement with the experimental gaps can be improved by increasing $\alpha$; we have determined that $\alpha(x)=0.2780+x\cdot0.0318$ produces band gaps of 1.00 eV for CIS and 1.72 eV for CGS, matching the experimental values. We have used the HSE06 functional with this $\alpha(x)$, the tuned HSE06 functional, in all computations presented in this article. The atomic positions in the bulk structure have been relaxed using a conjugate-gradient algorithm until all forces were below 0.01 eV/\AA. The calculated structural parameters ($a$, $c$ and the anion displacement $u$) are listed in Table \ref{tab:struc_par} and compared with experimental values \cite{Belhadj}. The alignment via the slabs is performed by means of the electrostatic potential (sum of the external potential due to the nuclei and the mean field electronic potential). In the slab the so-called macroscopic average electrostatic potential, \textit{i.e.}~the average of the planar average potential over distances of one unit cell along the transverse direction of the slab (\textit{cfr.}~Ref.~\citenum{Moses}), is computed. Using the fact that this macroscopic average in the middle of the slab should coincide with the average potential in the bulk material produces the desired alignment of the bands of the bulk structure with the potential in vacuum. In constructing the slab, the atomic positions of the bulk structure are kept fixed, so the potential in the center of the slab converges to that of the bulk material. The second method for band alignment requires that the number of valence and conduction bands to be used in the calculation of the $E_{BP}$ are specified. Guidelines for this choice have been established in Ref.~\citenum{Schleife}, which for the CIGS compounds with 64 electrons (in the $1 \times 1 \times 2$ supercell, not counting the d-electrons), lead to averaging over 16 valence bands and 8 conduction bands. 

\begin{table}[t]
\small
  \caption{\ The experimental structural parameters \cite{Belhadj} of CIS and CGS and structural parameters calculated with the tuned HSE06 functional. The relative deviation of the calculated values from the experimental values is added between parentheses.}
  \label{tab:struc_par}
  \begin{tabular*}{0.5\textwidth}{@{\extracolsep{\fill}}lllll}
    Parameter&CIS, exp.&CIS, calc.&CGS, exp.&CGS, calc. \\
    \hline
    $a$ (\AA)&5.784&5.832 &5.614&5.652 \\
    &&(+0.8\%)&& (+0.7\%)\\
    $c$ (\AA)&11.616&11.735 &11.030&11.119\\
    &&(+1.0\%)&& (+0.8\%) \\
    $u$ &0.224&0.229 &0.250&0.253 \\
    &&(+2.2\%)&&(+1.2\%) \\
    \hline
  \end{tabular*}
\end{table}

\subsection{Results and discussion}

\begin{figure}[h]
\centering
\includegraphics[height=5.5 cm]{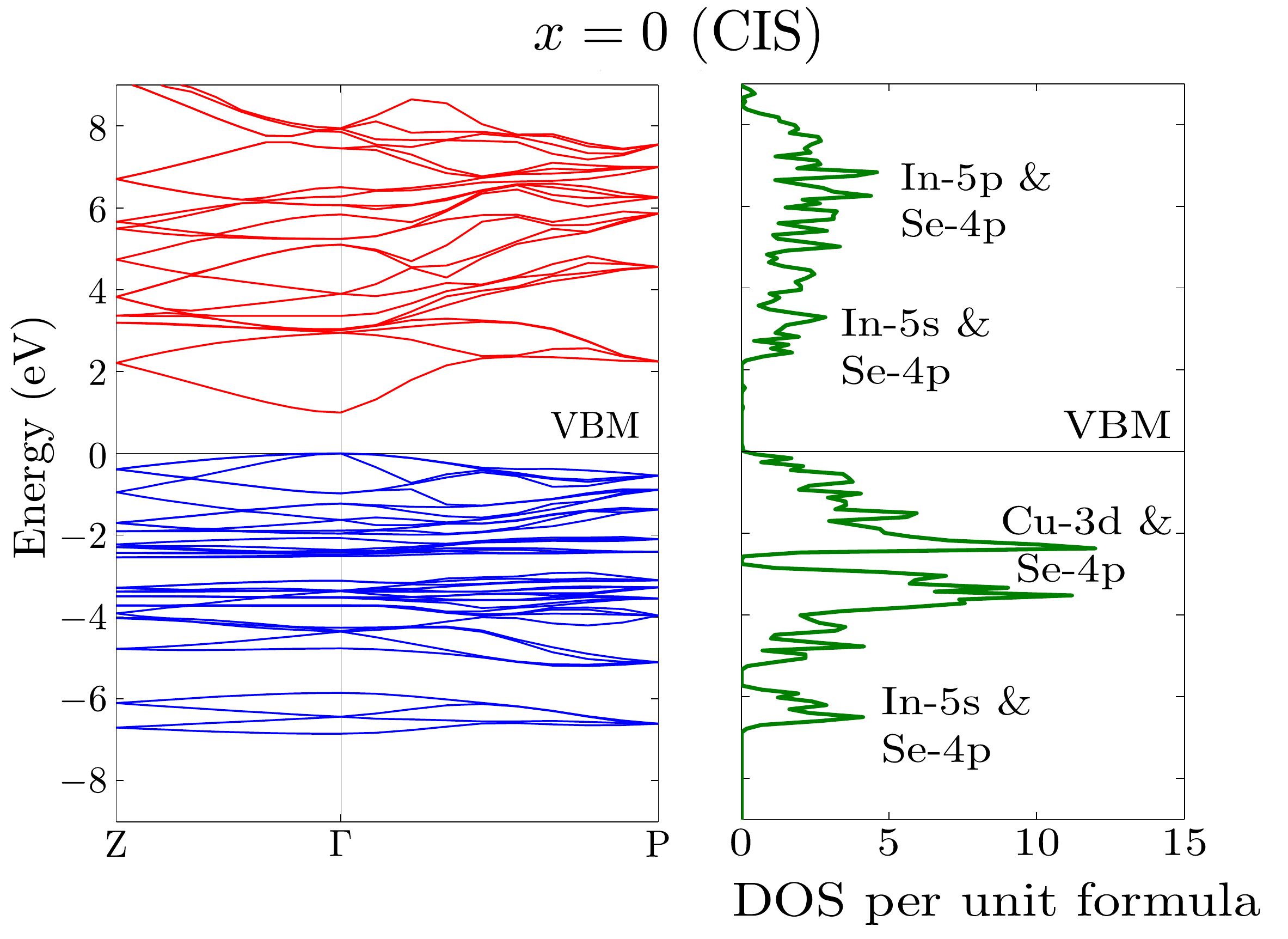}\llap{
  \parbox[b]{2.9in}{(a)\\\rule{0ex}{2.1in}
  }}

\vspace{0.3 cm}

\includegraphics[height=5.3 cm]{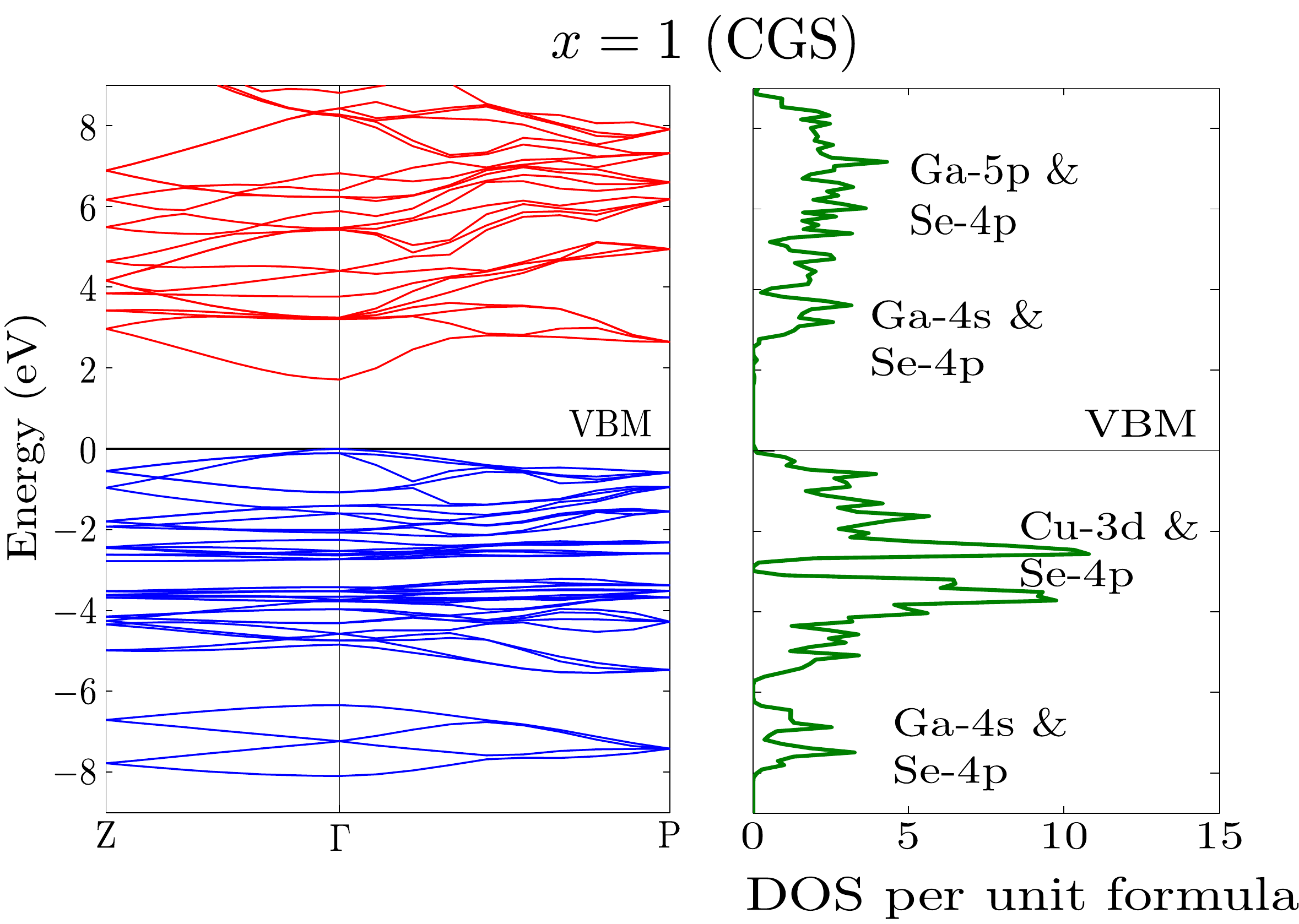}\llap{
  \parbox[b]{2.9in}{(b)\\\rule{0ex}{2in}
  }}
\caption{Band structure and density of states (DOS) per unit formula - containing 4 atoms - of (a) CIS and (b) CGS, obtained using the tuned HSE06 functional. The band structure is plotted along the Z-$\Gamma$-P path in reciprocal space \cite{Curtarolo}. The valence band maxima (VBM) are set to 0 eV. The character of the bands, as obtained from the study of the projected density of states, is added in the vicinity of the DOS.}
\label{fig:bands,dos}
\end{figure}
\begin{figure}[h]
\centering
\includegraphics[height=4 cm]{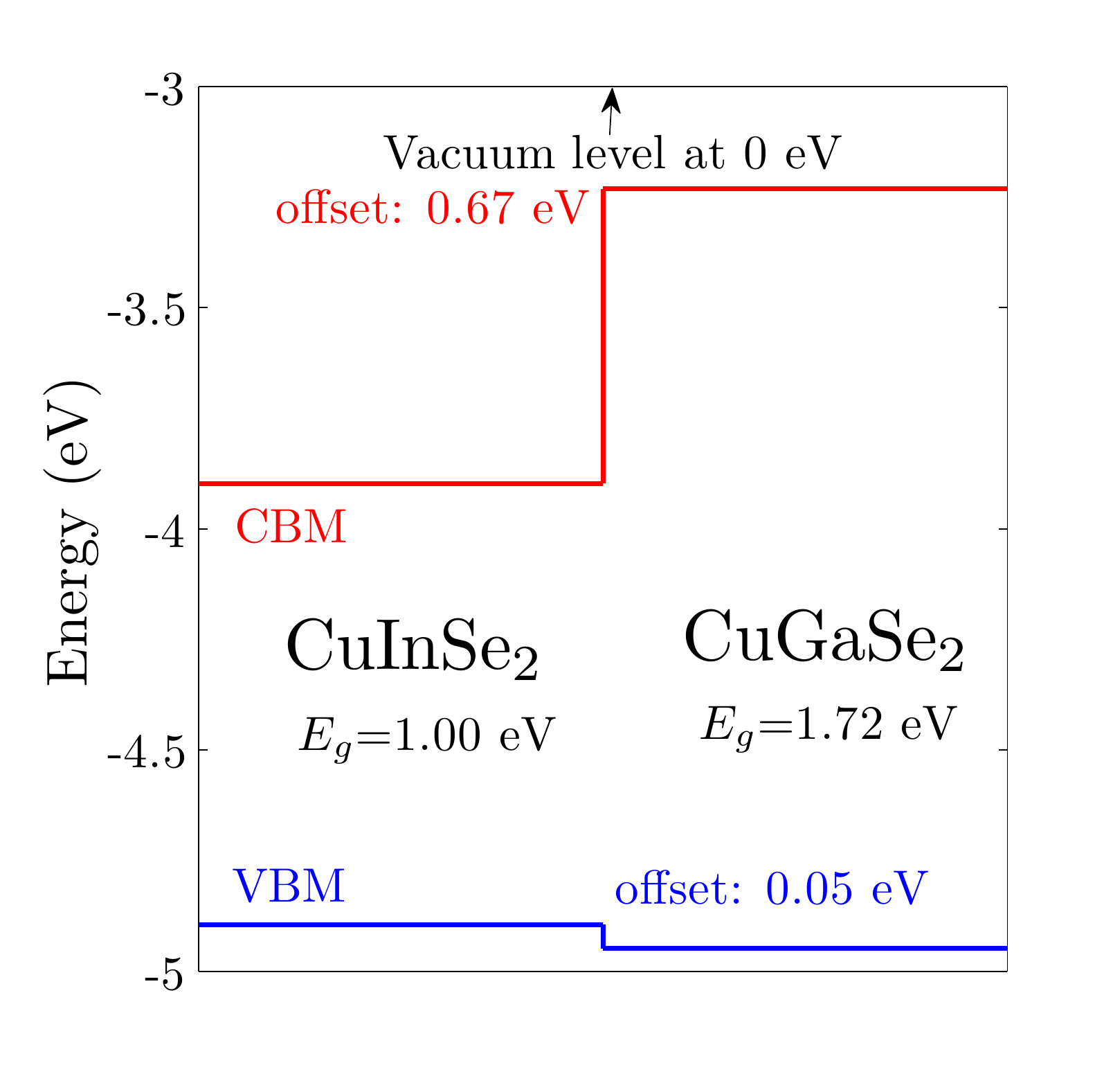}\llap{
  \parbox[b]{1.6in}{(a)\\\rule{0ex}{1.5in}
  }}
\includegraphics[height=3.98 cm]{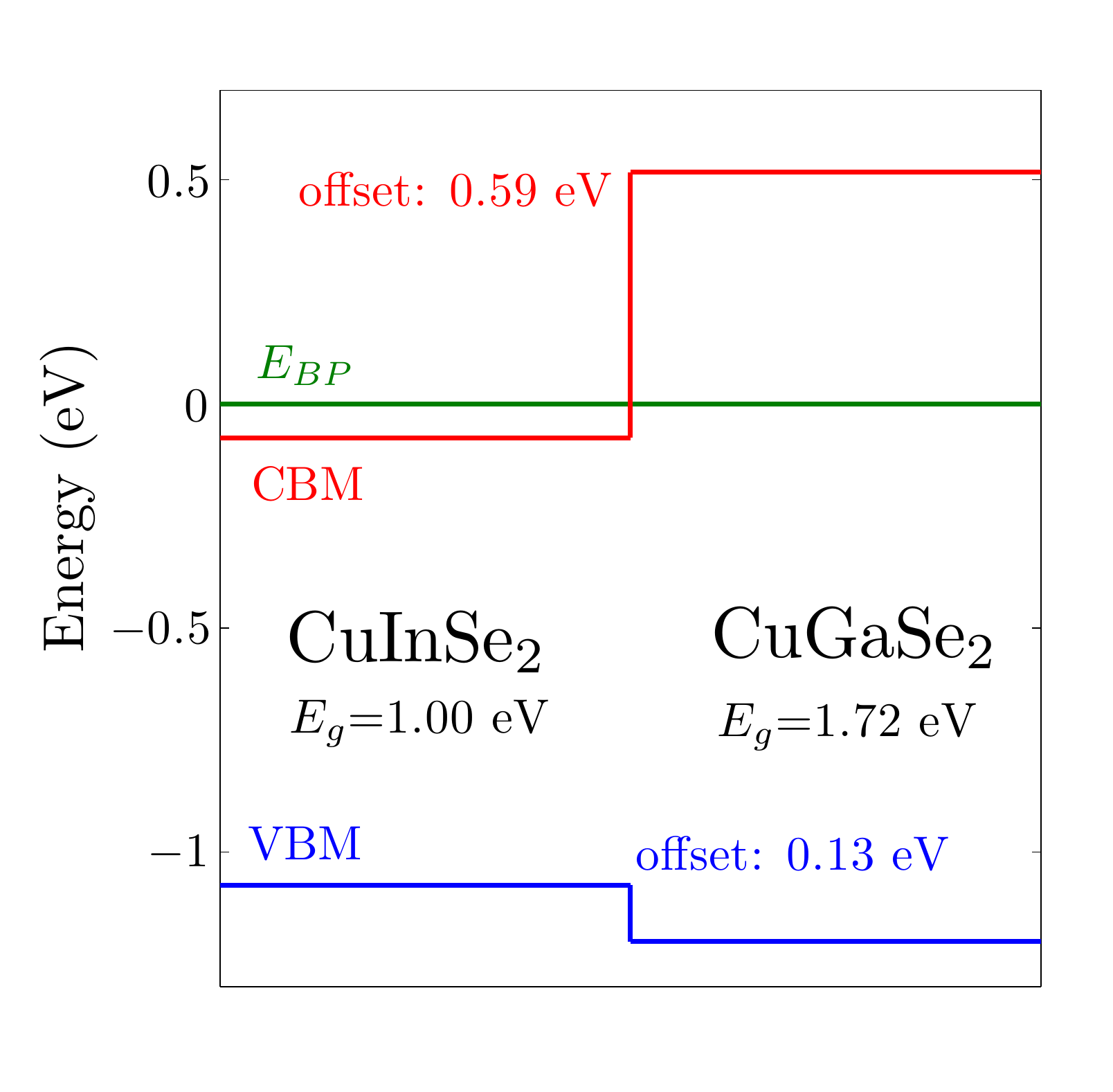}\llap{
  \parbox[b]{1.6in}{(b)\\\rule{0ex}{1.5in}
  }}

\vspace{0.1 cm}

\includegraphics[height=4 cm]{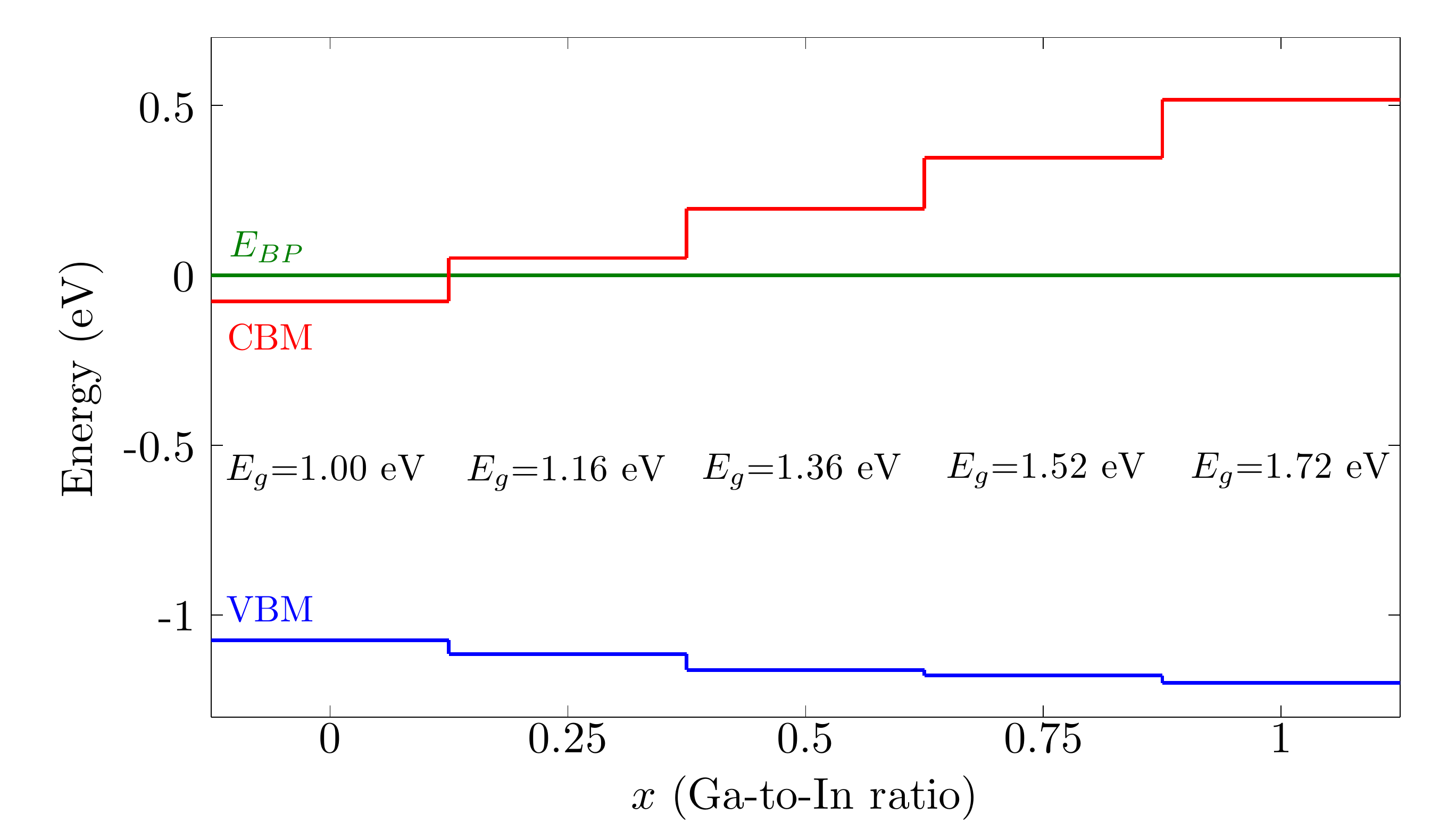}\llap{
  \parbox[b]{2.7in}{(c)\\\rule{0ex}{1.5in}
  }}
\caption{Comparison of the band alignment (VBM in blue, CBM in red) obtained by (a) slab calculations and (b) by using the $E_{BP}$, showing that both results are in good agreement. In (a) the vacuum potential level is set to 0 eV, while in (b) this is the case for the $E_{BP}$ (green line). The alignment method based on the $E_{BP}$ is extended to compounds with $x=0,~0.25,~0.5,~0.75,~1$ in (c). The band gaps ($E_g$) of the different compounds are also added to the plots.}
\label{fig:compar_alignments}
\end{figure}

In Fig.~\ref{fig:bands,dos}, the band structure and density of states (DOS) of CIS and CGS are displayed. An important property of both compounds is the clearly higher DOS near the valence band maximum (VBM) compared with the conduction band minimum (CBM). In this way, the Fermi level in pristine CIGS (at non-zero temperature) lies above midgap, thus facilitating p-type doping. The upper valence levels consist of hybridized Cu-3d and Se-4p states. As the strength of the hybridization interaction is inversely proportional to the energy separation of the p- and d-bands, the hybridization leads to so-called p-d repulsion, hereby inducing an upwards shift of the valence bands \cite{Jaffe}. The deeper lying valence levels shown in the figure as well as the lower conduction bands are primarily made up of In-5s/Ga-4s and Se-4p. The Se-4s and Ga-3d/In-4d states lie deeper in the band structure and are therefore not shown in the figure. The DOS is used further on in this article in the calculation of the net concentration of free charges.\\
\indent In Fig.~\ref{fig:bands,dos}, the VBM of CIS and CGS was quite arbitrarily set to 0 eV, so the band structures were not aligned. To accomplish the alignment, we compare the result using slabs and using the $E_{BP}$, first focussing on the limiting compounds, CIS and CGS. The results can be found in Fig.~\ref{fig:compar_alignments} (a) and (b). The two different methods for band alignment give consistent results, with a limited difference in offset of less than 0.1 eV between the VBM in CIS and that in CGS. In this way, the well-established alignment based on slabs justifies the application of the $E_{BP}$-based method to also treat intermediate $x=0.25,~0.5,~0.75$, as shown in Fig.~\ref{fig:compar_alignments} (c). The alignment between the adjacent compounds is of the so-called straddling type (type-I). Furthermore, the VBM alters much less with $x$ than the CBM. This is related to the character of the bands, namely upon replacing In with Ga the lower conduction bands containing In-5s character are naturally more affected than the upper valence bands mainly consisting of Cu-3d and Se-4p. Our result confirms - based on accurate calculations using the hybrid functional - the earlier band alignment of CIS and CGS by S.-H.~Wei and A.~Zunger \cite{Wei95}, obtained within the LDA, applying a rigid shift based on the experimental band gaps to treat the band gap problem. Finally, we wish to draw attention to the $E_{BP}$ itself. In compounds with low $x$, the $E_{BP}$ is found to be located close to the CBM, and it even lies within the conduction band in case of CIS. This means that there are donor-like surface states nearby the CBM, similar to \textit{e.g.}~in InN \cite{Mahboob}. It provides a possible explanation for the experimentally observed n-type conductivity at the surface of CIGS with low $x$, resulting in a type-inversion of the surface compared with the usually p-type interior \cite{Schmid}. In Ref.~\citenum{Schmid}, it is concluded - from the stoichiometry of the surface - that the type-inversion is due to the formation of the ordered defect compound CuIn$_3$Se$_5$. This proposed separate phase at the surface has however not been observed with direct methods such as X-ray diffraction, moreover its effect has been estimated to be insufficient to account for the observed type-inversion \cite{Markvart}. Instead, other models have been proposed to explain the type-inversion, including donor defects due to dangling bonds \cite{Cahen} and a barrier for holes due to surface reconstruction \cite{Persson2003}. In addition to, or as an alternative for these proposed models, our calculation of the $E_{BP}$ show n-type behavior at the CIS surface, independent of the structural details of the surface. 

\section{Native point defects}

As we have discussed in the Introduction, photoluminescence spectra show that the cation (Cu, In, Ga) stoichiometry plays a major role in the properties of the native defects in undoped CIGS. This is our rationale to compute the point defects related to a varying stoichiometry of the cations, in CIS and CGS. Hence, for CIS (CGS) we compute the vacancy defects V$_{\mathrm{Cu}}$ (idem) and V$_{\mathrm{In}}$ (V$_{\mathrm{Ga}}$) and the antisite defects Cu$_{\mathrm{In}}$ (Cu$_{\mathrm{Ga}}$) and In$_{\mathrm{Cu}}$ (Ga$_{\mathrm{Cu}}$). We proceed as follows: first we briefly review the formalism needed to calculate the formation energy of point defects from first-principles (see also e.g.~Ref.~\citenum{Persson2005,Wei1998}), thereby elaborating on the important role of the chemical potentials of the exchanged atoms. Then, we sketch how this enables us to determine the Fermi level self-consistently, from overall charge neutrality. This is used to estimate the conductivity type and free charge concentration depending on growth conditions.

\subsection{Method}

\subsubsection{Formation energy, transition levels.~~}

The formation energy of defect $\mathcal{D}$ in charge state $q$, $E_f\left(\mathcal{D},q\right)$, a Gibbs free energy, is defined as
\begin{align}
E_f\left(\mathcal{D},q\right)=E_{\mathrm{tot}}\left(\mathcal{D},q\right)-E_{\mathrm{tot}}\left(\mathrm{bulk}\right)+\sum_{\nu}n_{\nu}\mu_{\nu} 
\notag \\ 
+q\left(E_{VBM}+E_F+\Delta V^{(q)}\right)~.
\label{eq:eformation}
\end{align}
In this expression, $E_{\mathrm{tot}}\left(\mathcal{D},q\right)$ is the total energy of the supercell containing the defect and $E_{\mathrm{tot}}\left(\mathrm{bulk}\right)$ is the total energy of the bulk supercell (\textit{i.e.}~without defect). In the third term, $\mu_{\nu}$ are the chemical potentials of the exchanged atoms. The absolute value $\left|n_{\nu}\right|$ is the number of exchanged atoms of element $\nu$; furthermore if the atoms are added $n_{\nu}<0$, in case they are removed $n_{\nu}>0$. For charged defects, the last term takes into account the exchange of electrons ($q<0$ if they are added to the supercell and $q>0$ if they are removed) with the Fermi level $E_F$, referenced to $E_{VBM}$, the top of the valence band of the bulk cell. Finally, $\Delta V^{(q)}$ is the difference in reference potential of the supercell without defect and with defect. J.~L.~Lyons \textit{et al.}~have studied several corrections for the effects of the finite size supercell and have reported the correction scheme based on $\Delta V^{(q)}$ to be consistent with other correction schemes, such as the Madelung correction \cite{Lyons2012}. From Eq.~\ref{eq:eformation}, it follows thus that the formation energies of the defects are linear functions of $E_F$. The Fermi level at which the formation energy functions of different charge states $q$ and $q'$ of a certain defect intersect, is called the transition level $\varepsilon\left(\mathcal{D},q/q'\right)$. From Eq.~\ref{eq:eformation} it follows this transition level can be calculated as
\begin{align}
\varepsilon(\mathcal{D},q/q')=&\frac{E_{tot}(\mathcal{D},q)-E_{tot}(\mathcal{D},q')+q\Delta V^{(q)}- q' \Delta V^{(q')}}{q'-q}\notag \\&\hspace{4.5cm}-E_{VBM} ~.
\label{eq:trans_level}
\end{align}
The transition levels relative to the valence and conduction band determine the electrical activity of the defect state. 

\subsubsection{Chemical potential range.~~}

\begin{figure}[t]
\centering
\includegraphics[height=5 cm]{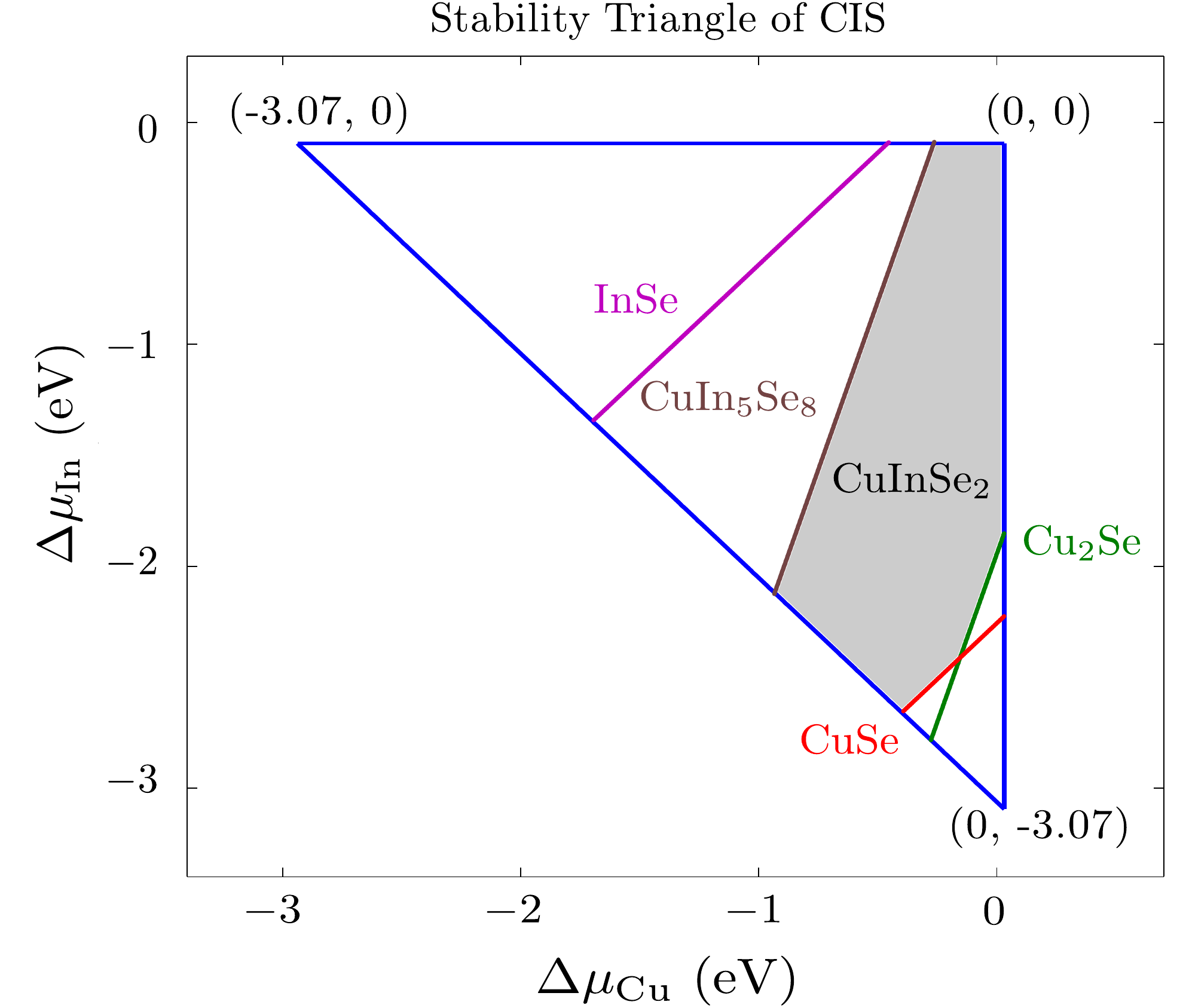}\llap{
  \parbox[b]{2.5in}{(a)\\\rule{0ex}{1.8in}
  }}
\vspace{0.5 cm}

\includegraphics[height=5.5 cm]{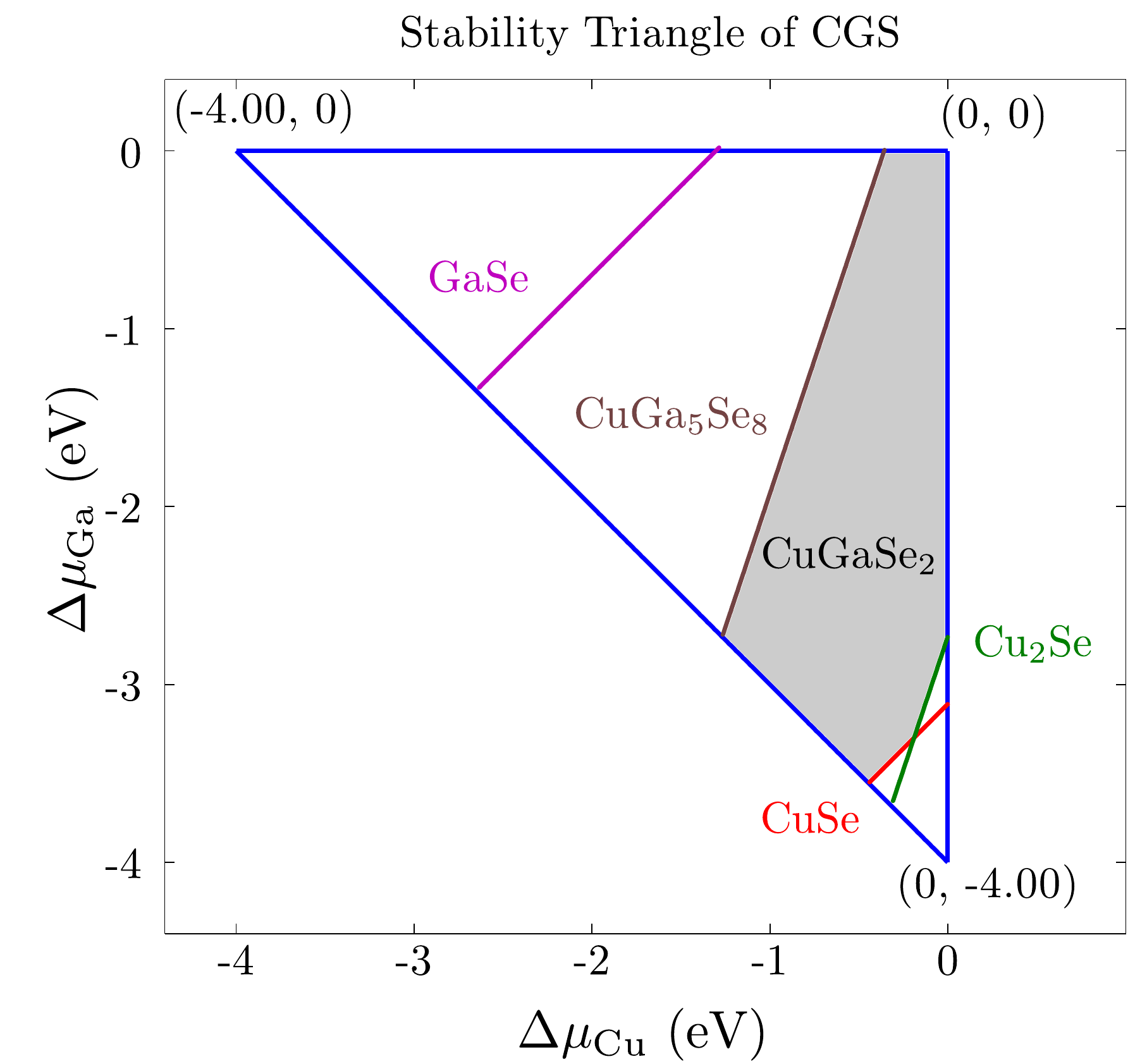}\llap{
  \parbox[b]{2.5in}{(b)\\\rule{0ex}{1.9in}
  }}
\caption{The stability triangles of (a) CIS and (b) CGS as a function of $\Delta \mu_{Cu}$ and $\Delta \mu_{In/Ga}$. Within the triangles, the gray areas represent the stable chemical potential range of CIS and CGS following from the restriction by the competing phases.}
\label{fig:triangles}
\end{figure}

The chemical potential $\mu_{\nu}$ of element $\nu$ in the crystal is the free energy of the atoms of this element in the reservoir in contact with the system. As such, the chemical potentials depend on the experimental growth conditions. The chemical potential can be rewritten as the sum of the chemical potential of the elemental phase ($\mu_{\nu}^{\mathrm{elem}}$) and a deviation $\Delta \mu_{\nu}$, where a more negative $\Delta \mu_{\nu}$ means $\nu$-poorer growth conditions. In thermodynamic equilibrium, the deviations are subject to three constraints \cite{Persson2005}. 
\begin{enumerate}
	\item In order to avoid precipitation of the elemental phase, all $\mu_{\nu}\leq \mu_{\nu}^{\mathrm{elem}}$; this means for \textit{e.g.}~CIS: $\Delta \mu_{\mathrm{Cu}}\leq 0~, ~\Delta \mu_{\mathrm{In}}\leq 0~,~\Delta \mu_{\mathrm{Se}}\leq 0$. 
	\item The formation of a stable compound requires that the sum of the $\Delta \mu_{\nu}$ equals the heat of formation $\Delta H_f$, \textit{i.e.}~the difference of total energy of a compound and the energy of the constituent atoms in their elemental phase. For CIS this translates to: $\Delta H_f(\mathrm{CIS})=\Delta \mu_{\mathrm{Cu}}+\Delta \mu_{\mathrm{In}}+2\Delta \mu_{\mathrm{Se}}$.
	\item The formation of competing phases also lays a restriction on the accessible chemical potential range. In case of CIS (CGS), the competing phases we take into account are Cu$_2$Se (idem), CuSe (idem), InSe (GaSe) and the ordered defect compound CuIn$_5$Se$_8$ (CuGa$_5$Se$_8$). The respective space groups of these compounds are Fm$\bar{3}$m (cubic), P$6_{3}$/mmc (hexagonal), P$6_{3}$/mmc (hexagonal) and P$\bar{4}$ (tetragonal) \cite{Pearson,Duran}. As an example, the competition of Cu$_2$Se leads to the constraint $2\Delta \mu_{\mathrm{Cu}}+\Delta \mu_{\mathrm{Se}} \leq \Delta H_f(\mathrm{Cu}_2\mathrm{Se})$.
\end{enumerate}
By applying these constraints we can construct the accessible chemical potential range, for a ternary compound the so-called called the stability triangle, shown in Fig.~\ref{fig:triangles} as a function of $\Delta \mu_{\mathrm{Cu}}$ and $\Delta \mu_{\mathrm{In/Ga}}$. The third chemical potential, $\Delta \mu_{\mathrm{Se}}$, is a dependent variable, as follows from constraint No.~2. All calculations used to obtain the stability triangles have been performed with the HSE06 functional. It can be observed in Fig.~\ref{fig:triangles} that the chemical range for which CIS and CGS are formed is quite broad. This characteristic of CIGS has also been confirmed experimentally \cite{Godecke}. Overall, we see many similarities with previous theoretical results, \textit{e.g.}~in Ref.~\citenum{Persson2005,Wei2013}. But, compared with Ref.~\citenum{Persson2005,Wei2013}, we obtain a better agreement of $\Delta H_f(\mathrm{CIS})$ and $\Delta H_f(\mathrm{CGS})$ with experiment, due to the accurate calculation of total energies with the HSE06 functional. Theoretically, we find $\Delta H_f(\mathrm{CIS})=-3.07$ eV and $\Delta H_f(\mathrm{CGS})=-4.00$ eV, consistent with the experimental values of $-2.77$ eV and $-3.29$ eV respectively \cite{Verma}. A second important difference with Ref.~\citenum{Wei2013} is our result that InSe and GaSe do not put an extra restriction on the formation of CIS and CGS, according to Fig.~\ref{fig:triangles}. For these compounds we can likewise demonstrate a good correspondence between our results $\Delta H_f(\mathrm{InSe})=-1.38$ eV and $\Delta H_f(\mathrm{GaSe})=-1.28$ eV and the experimental values $-1.63$ eV and $-1.10$ eV respectively \cite{Colin,Mills}. 

\subsubsection{Self-consistent determination of the Fermi level.~~}

In semiconductor systems, in which both donors and acceptors are present and have similar formation energies, it is important to determine the position of the Fermi level self-consistently, from charge neutrality (\textit{i.e.}~conservation of total charge). This charge neutrality is expressed by the balance $p+N_D^*=n+N_A^*$ between the hole and electron concentrations $p$ and $n$ and the concentrations of excess charges of ionized donors and acceptors $N_{D}^*$ and $N_{A}^*$. The hole and electron concentrations follow from the following integrals of the product of the DOS, $D(E)$, and the Fermi-Dirac distribution
\begin{align}
p=\int\limits_{-\infty}^{E_{VBM}}D(E)\frac{1}{1+\mathrm{exp}\left[(E_F-E)/(k_B T)\right]} dE~,\notag\\
n=\int\limits_{E_{CBM}}^{+\infty}D(E)\frac{1}{1+\mathrm{exp}\left[(E-E_F)/(k_B T)\right]}dE~,
\label{eq:chcarrconcentr}
\end{align}
where $E_{VBM}$ and $E_{CBM}$ are the VBM and CBM. We solve these integrals numerically, using the DOS shown in Fig.~\ref{fig:bands,dos}.
In thermodynamic equilibrium, the charged defect concentrations contributing to $N_{D}^*$ and $N_{A}^*$ follow a Boltzmann distribution depending on the defect formation energies (as can be found in many references, including Ref.~\citenum{VandeWalle1993,Ma2011}). Therefore, the concentration of defect $\mathcal{D}$ in charge state $q$ is given by
\begin{align}
N(\mathcal{D},q)=g_q~M_{\mathcal{D}}~\mathrm{exp}\left[-E_f(\mathcal{D},q)/(k_B T)\right]~,
\label{eq:defconcentr}
\end{align}
where $M_{\mathcal{D}}$ denotes the lattice site multiplicity where the defect can originate and $g_{q}$ is a degeneracy factor for charge state $q$. This factor depends on the electronic degeneracy, including spin degeneracy \cite{Ma2011}. The electronic degeneracies can be obtained as follows, by investigating the levels due to the defect. In creating V$_{\mathrm{Cu}}$, the doubly degenerate defect level due to the broken bond contains one electron and one hole in the neutral case. Therefore, the number of possible electronic configurations is given by the combination {\tiny $\left( \begin{array}{c} 2 \\ n \end{array} \right)$}, where $n$ represents the number of electrons. So, for $q=0$ ($n=1$) one obtains $g_0=2$, while $q=-1$ ($n=2$) is non-degenerate with $g_{-1}=1$. Similarly, the defect levels due to V$_{\mathrm{In/Ga}}$ are six-fold degenerate and contain 3 electrons and 3 holes, at $q=0$. The different combinations {\tiny $\left( \begin{array}{c} 6 \\n \end{array} \right)$} with $n=3,4,5,6$ lead to degeneracies $g_0=20$, $g_{-1}=15$, $g_{-2}=6$ and $g_{-3}=1$. The levels of the substitutional defect Cu$_{\mathrm{In/Ga}}$ are also six-fold degenerate and contain 4 electrons and 2 holes. This gives rise to combinations {\tiny $\left( \begin{array}{c} 6 \\n \end{array} \right)$} with $n=4,5,6$ and as such $g_0=15$, $g_{-1}=6$ and $g_{-2}=1$. Finally, the substitutional defect In/Ga$_{\mathrm{Cu}}$ forms a doubly degenerate defect level filled by 2 electrons, yielding degeneracies $g_0=1$, $g_{+1}=2$ and $g_{+2}=1$. To calculate the concentration of excess charges due to donors and acceptors $N_{D}^*$ and $N_{A}^*$ for charge neutrality from Eq.~\ref{eq:defconcentr}, the defect concentrations have to be multiplied by $\vert q \vert$ and summed over the different charges $q$. Eq.~\ref{eq:defconcentr} holds for donors and acceptors, both shallow and deep and it produces some well-known formulas for simple cases, such as the concentrations of singly ionizable donors and acceptors \cite{Sze,Chin}. To achieve this, one writes out the fraction of ionized acceptors $N(\mathcal{D},q=-1)/N_A$ with $N_A=N(\mathcal{D},q=0)+N(\mathcal{D},q=-1)$, the total number of these acceptors. Subsequently, one uses that $E_f(\mathcal{D},q=-1)-E_f(\mathcal{D},q=0)=\epsilon(0/-1)-E_F$, as follows from Eqs.~\ref{eq:eformation} and \ref{eq:trans_level}. The result is that the concentration of defects $\mathcal{D}$ ionized to charge state $q=-1$, for instance V$_{\mathrm{Cu}}$, is given by the following product of $N_A$ and a Fermi-Dirac-like distribution:
\begin{align}
N(\mathcal{D},q=-1)=\frac{N_{A}}{1+\frac{g_0}{g_{-1}}~\mathrm{exp}\left[(\epsilon(0/-1)-E_F)/(k_B T)\right]}~,
\label{eq:FDdefects}
\end{align}
and in case of V$_{\mathrm{Cu}}$, $g_0/g_{-1}=2$. A similar expression can be derived for donors, for which the Fermi-Dirac-like distribution depends on $(E_F-\epsilon(+1/0))$. To simulate the experiments more closely, we calculate the total concentration of a defect - as a sum of the concentrations due to the different charge states - at $800$ K. This is the temperature around which CIGS samples are usually grown, in coevaporation and in vacuum-based sequential growth methods, during selenization \cite{Singh}. It is likely that the total concentration formed during growth freezes in during cooling down, due to kinetic barriers \cite{Persson2005}. The ratios between the different charge states of the defects are subsequently calculated via the Boltzmann distribution at $300$ K (the temperature at which the photovoltaic device is operated) and the concentrations per charge state redistributed accordingly. Notice that the formation energy determining the defect concentrations (Eqs.~\ref{eq:eformation} and \ref{eq:defconcentr}) is a function of $E_F$, while at the same time $E_F$ is determined by the defect concentrations through charge neutrality. Hence, $E_F$ is to be found self-consistently, which we denote $E_F^{SC}$; we have performed this task numerically. In this approach, a problem arises when $E_f(\mathcal{D},q)$ becomes negative, also known as spontaneous formation. To avoid the defect concentration following from Eq.~\ref{eq:defconcentr} exceeding the lattice multiplicity, we limit the defect concentrations to the available number of lattice points. In light of this, the method we have presented does not allow to predict the absolute defect concentrations. Nevertheless, $E_F^{SC}$ is almost invariant of a uniform (\textit{i.e.}~for all defects) shift of the formation energies that limits the formation. This shows that first and foremost the \emph{ratios} of the defect concentrations, depending on their formation energies, matter in the determination of $E_F^{SC}$. As such, our calculations lead to reasonable values for $E_F^{SC}$, from which we determine $p-n$, the net concentration of free charges. If $p-n>0$, the material is p-type, else if $p-n<0$, it is n-type. Equivalently, for p-type $E_F^{SC}<E_F^i$, where by $E_F^i$ we mean the intrinsic Fermi level. It is the Fermi level in the pristine material corresponding to charge neutrality $p=n$. At nonzero temperatures it is not necessarily located at midgap, if the DOS of the valence and conduction bands are different. Since in CIGS - as we have mentioned when discussing Fig.~\ref{fig:bands,dos} - the DOS near the VBM is higher than near the CBM, $E_F^i$ lies above midgap. We find that $E_F^i=0.57$ eV for CIS (midgap: 0.50 eV) and $E_F^i=0.90$ eV for CGS (midgap: 0.86 eV), at a temperature of 300 K.

\subsection{Computational details}

All results have been obtained with the tuned HSE06 functional, discussed in the computational details of the band alignment. In fact, many of the computational details are unchanged, yet a few techniques are specific to point defect calculations. The defects are positioned in a supercell of the primitive cell, to minimize the electrostatic interactions between the defects, thereby also minimizing band filling. These contributions would add to the total energy needed to calculate the formation energy. To test this, we have compared the formation energies of V$_{\mathrm{Cu}}$ in CIS and CGS in a $2\times 2 \times 2$ supercell containg 64 atoms and a $3\times 3 \times 3$ supercell containing 216 atoms and found that the former already yields well-converged values. Thus, we present formation energies in the 64-atom supercell. The atomic positions in the supercells containing a defect are relaxed until all forces are smaller than 0.05 eV/\AA, keeping the volume of the cell fixed. Due to the dimensions of the cell, the \textbf{k}-point grid used for integration over the Brillouin zone scales to $2 \times 2\times 2$. The charge state $q$ of the defect is simulated by adding $q=.., -2, -1, 0, +1, +2, ..$ electrons to the supercell. We calculate the correction for the reference potential $\Delta V^{(q)}$ (Eq.~\ref{eq:eformation}) as described in Ref.~\citenum{Amini}.

\subsection{Results and discussion}

\begin{figure}[t]
\centering

\hspace{0.1 cm}

\includegraphics[height=3.95 cm]{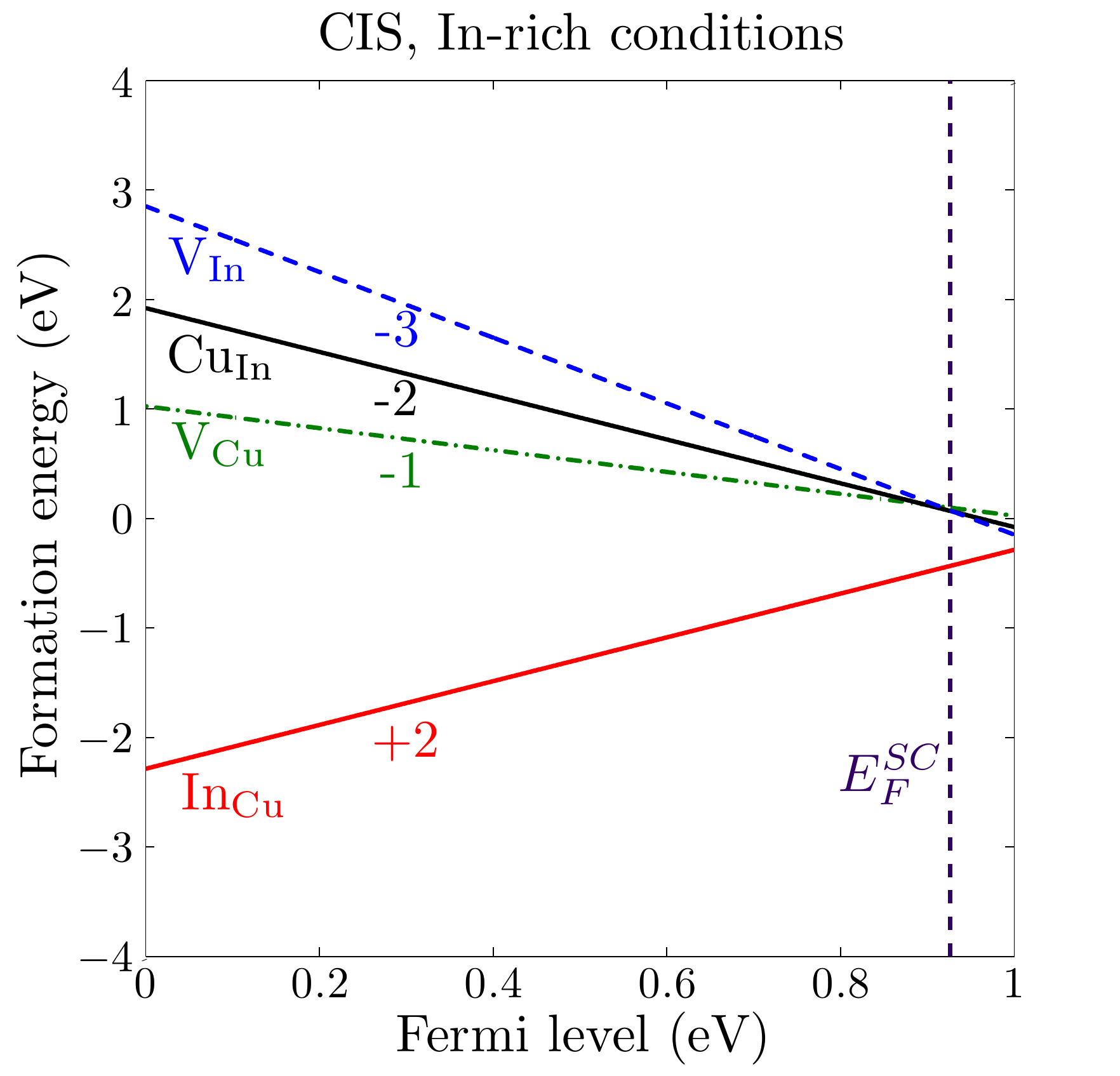}\llap{
  \parbox[b]{1.6in}{(a)\\\rule{0ex}{1.5in}
  }}
\includegraphics[height=3.9 cm]{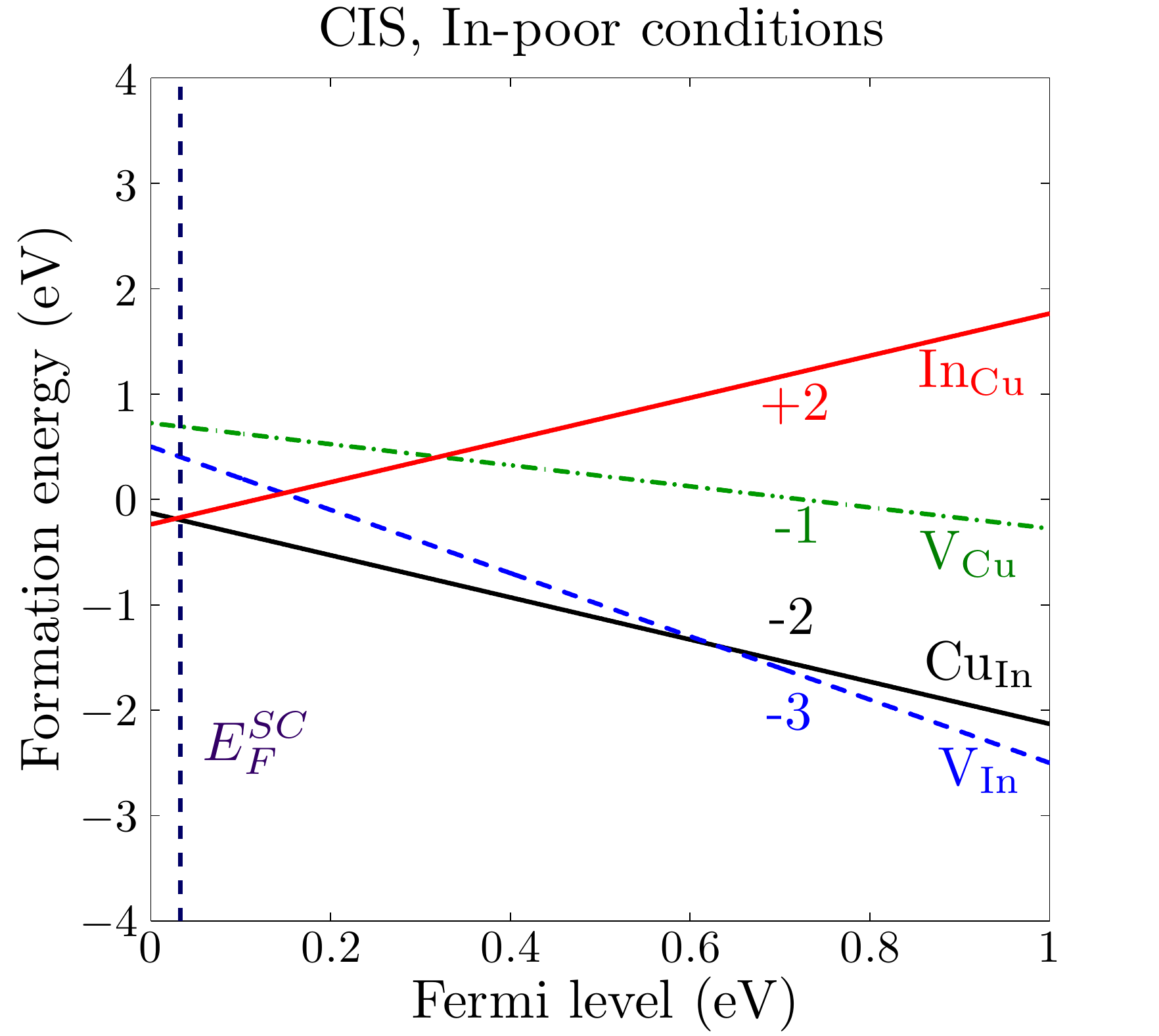}\llap{
  \parbox[b]{1.6in}{(b)\\\rule{0ex}{1.5in}
  }}

\vspace{0.25cm}

\includegraphics[height=3.85 cm]{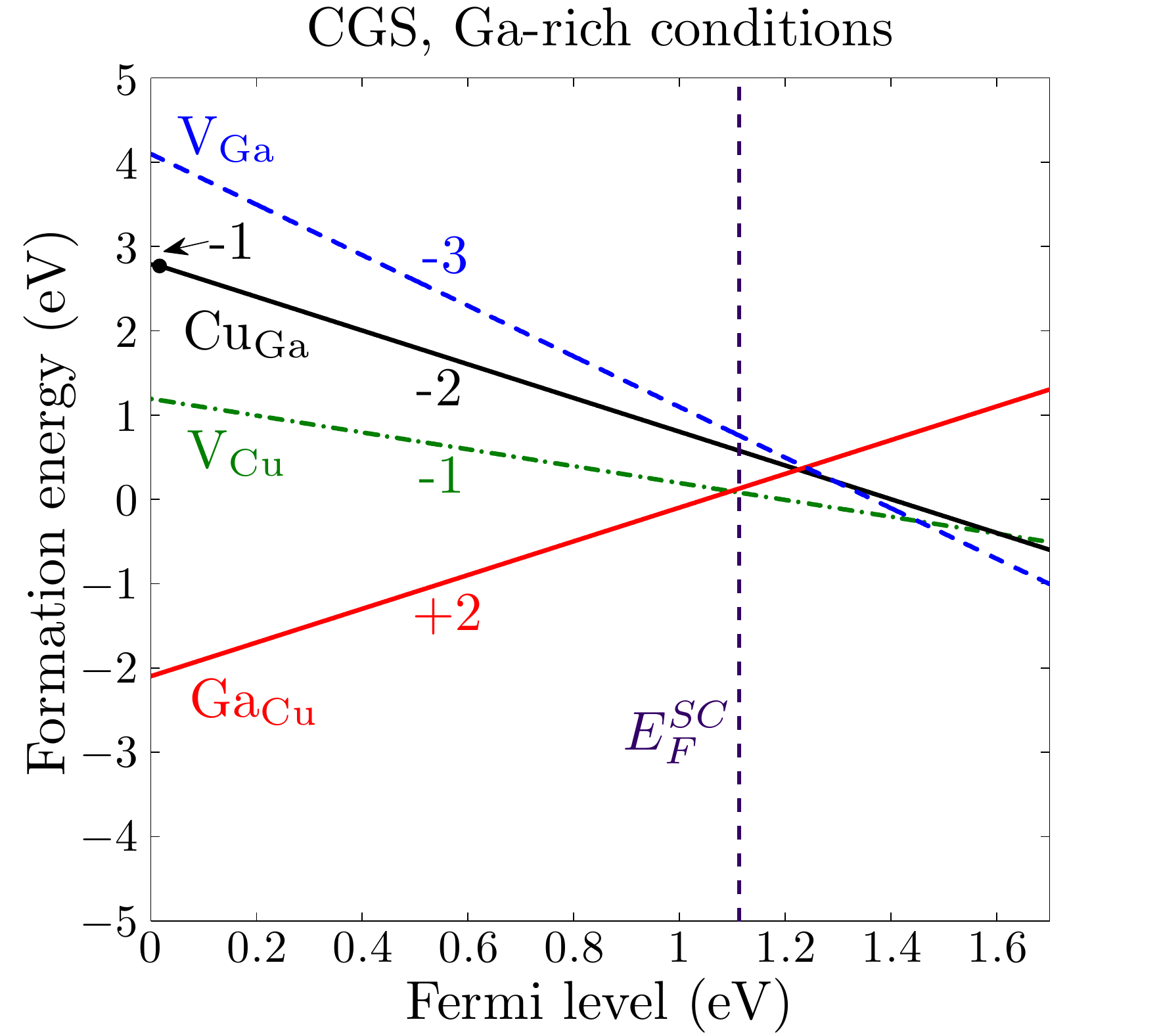}\llap{
  \parbox[b]{1.7in}{(c)\\\rule{0ex}{1.5in}
  }}
\includegraphics[height=3.95 cm]{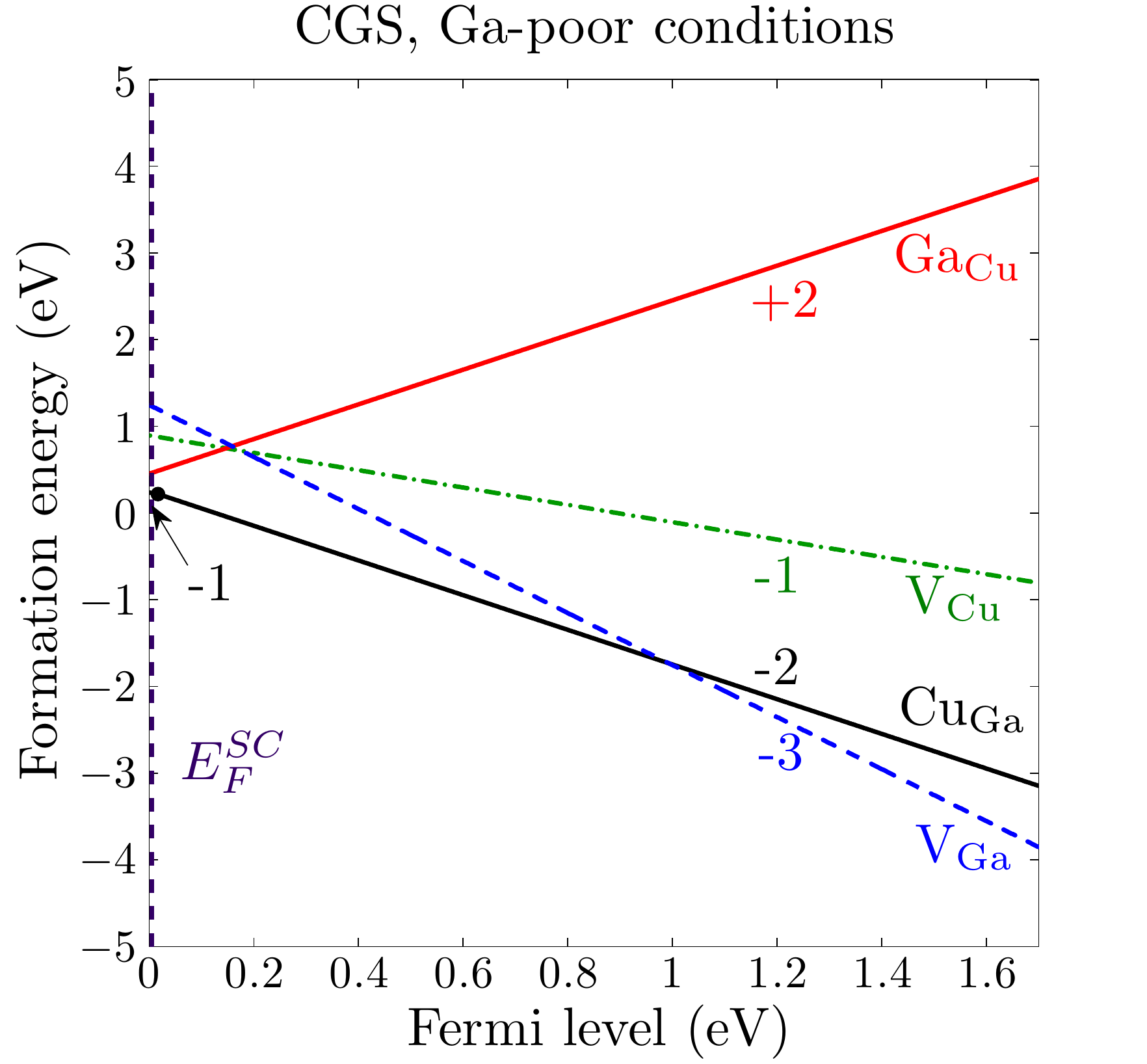}\llap{
  \parbox[b]{1.6in}{(d)\\\rule{0ex}{1.5in}
  }}
\caption{Formation energies (eV) of ground charge states of V$_{\mathrm{Cu}}$ (idem), V$_{\mathrm{In}}$ (V$_{\mathrm{Ga}}$), Cu$_{\mathrm{In}}$ (Cu$_{\mathrm{Ga}}$) and In$_{\mathrm{Cu}}$ (Ga$_{\mathrm{Cu}}$) in CIS (CGS), as function of the Fermi level between VBM and CBM. The charge states are listed near the curves, while the transition levels are indicated by solid dots. $E_F^{SC}$, represented by a dashed vertical line, is determined through charge neutrality. For CIS, we distinguish between (a) In-rich conditions $\left(\Delta \mu_{\mathrm{Cu}},\Delta \mu_{\mathrm{In}}\right)=\left(-0.2,-0.15\right)$ eV and (b) In-poor conditions $\left(\Delta \mu_{\mathrm{Cu}},\Delta \mu_{\mathrm{In}}\right)=\left(-0.5,-2.5\right)$ eV. Similarly, for CGS we compare (c) Ga-rich conditions $\left(\Delta \mu_{\mathrm{Cu}},\Delta \mu_{\mathrm{Ga}}\right)=\left(-0.2,-0.15\right)$ eV with (d) Ga-poor conditions $\left(\Delta \mu_{\mathrm{Cu}},\Delta \mu_{\mathrm{Ga}}\right)=\left(-0.5,-3.0\right)$ eV.}
\label{fig:form_energies}
\end{figure}
Within the hybrid functional method, we have obtained the formation energies shown in Fig.~\ref{fig:form_energies}, in which only the formation of the ground charge state (with lowest formation energy) is plotted. The formation energies are a function of $E_F$, which is referenced to the VBM, \textit{i.e.}~$E_F=0$ corresponds to the VBM and $E_F=E_g$ to the CBM. As we have established before, the formation energies depend on the chemical potentials of the constituent elements, within an allowed range (shown in Ref.~\ref{fig:triangles}). In CIS and CGS, the two most distinct regimes are In/Ga-rich and In/Ga-poor. Therefore, we initially focus on a few characteristic examples, afterwards discussing the conductive properties in the entire existence region of the host materials. The transition levels between the charge states of a specific defect are of course not dependent on the chemical potentials. In case of In$_{\mathrm{Cu}}$ and Ga$_{\mathrm{Cu}}$, the $q=+2$ charge state has the lowest formation energy, therefore these defects donate 2 electrons. The transition to the neutral charge states ($q=0$) takes place at $E_F>E_g$ (referenced to the VBM), so In$_{\mathrm{Cu}}$ and Ga$_{\mathrm{Cu}}$ act as shallow donors. The other defects act as acceptors, since the negative charge state has the lowest formation energy. The vacancies V$_{\mathrm{Cu}}$ and V$_{\mathrm{In/Ga}}$ prefer to be in charge states $q=-1$ and $q=-3$, respectively. The antisite defect Cu$_{\mathrm{In}}$ in CIS has $q=-2$ as its ground charge state for all $E_F$ in the band gap, while its equivalent in CGS, Cu$_{\mathrm{Ga}}$, has a very shallow transition $\varepsilon\left(\mathrm{Cu}_{\mathrm{Ga}},-1/-2\right)=0.017$ eV. All acceptor states are shallow, since the transition to $q=0$ occurs for $E_F<0$.\\ 
\indent In general, there are significant differences between our results and previous results by other authors, obtained by calculations based on the LDA. There is a consensus on V$_{\mathrm{Cu}}$ between Ref.~\citenum{Persson2005,Wei1998,Wei2013} and our results, but the other acceptor-type defects V$_{\mathrm{In/Ga}}$ and Cu$_{\mathrm{In}/\mathrm{Ga}}$ are predicted to be deep in Ref.~\citenum{Wei1998,Wei2013}. Similarly, the donor-type defects In/Ga$_{\mathrm{Cu}}$ are reported to be deep in both Ref.~\citenum{Persson2005} and Ref.~\citenum{Wei1998,Wei2013} and in the latter the formation energy of these defects is also quite high. We expect that these differences show the limitations of LDA both in predicting the defect formation energies and the transition levels, which have been thoroughly discussed in Ref.~\citenum{Janotti}. Our results have in general closer agreement with the hybrid functional results in Ref.~\citenum{Pohl2013}. However, in the latter the authors report that Cu$_{\mathrm{In}/\mathrm{Ga}}$ are deep acceptors, again raising the question how the p-type conductivity is maintained and even increases in Cu-rich samples. Another point of disagreement is their considerably higher formation energy of V$_{\mathrm{In/Ga}}$ compared with our values. As a consequence, they expect only two acceptor levels to be detected in experiment, whereas in several experiments three acceptors are found, as we have mentioned in the Introduction.\\
\indent Since there are both shallow acceptors and donors - with low formation energy - present in CIGS, the chemical potentials of the elements (the growth conditions) play a determining role. Under increasingly In/Ga-rich conditions (or equivalently Se-poor as $\Delta \mu_{\mathrm{Cu}}$ also becomes small due the chemical potential range presented in Fig.~\ref{fig:triangles}), the formation energy of the donor In/Ga$_{\mathrm{Cu}}$ decreases. As a result, the Fermi level following from charge neutrality is pushed above $E_F^i$. An example for CIS in this regime is depicted in Fig.~\ref{fig:form_energies} (a), where $E_F^{SC}=0.93$ eV, yielding strongly n-type conditions, with net concentration of electrons of $n-p=1.9\cdot 10^{17}$ cm$^{-3}$ at 300 K. In In/Ga-poor conditions the formation energy of In/Ga$_{\mathrm{Cu}}$ rises, while the formation energy of the acceptors V$_{\mathrm{In/Ga}}$ and Cu$_{\mathrm{In/Ga}}$ lowers, enhancing the formation of these acceptors. Maximum p-type conditions are thus, within the attainable chemical potential range in CIS, met around $\left(\Delta \mu_{\mathrm{Cu}},\Delta \mu_{\mathrm{In}}\right)=\left(-0.5,-2.5\right)$ eV. These growth conditions yield the formation energies plotted in Fig.~\ref{fig:form_energies} (b). The resulting self-consistently determined Fermi level is $E_F^{SC}=0.033$ eV, with a net concentration of holes amounting to $p-n=9.8\cdot 10^{19}$ cm$^{-3}$ at 300 K. In contrast with CIS, n-type conductivity in CGS is limited by the acceptor-type defects. They require significantly less formation energy if $E_F$ lies above midgap in the wider-gap compound ($E_g=1.72$ eV), thus pinning $E_F^{SC}$ far from the CBM. We find that in the Ga-rich conditions in Fig.~\ref{fig:form_energies} (c) $E_F^{SC}=1.1$ eV, resulting in a free electron concentration of merely $n-p=1.9\cdot 10^{9}$ cm$^{-3}$. Similarly to CIS, maximum p-type conditions in CGS are reached around $\left(\Delta \mu_{\mathrm{Cu}},\Delta \mu_{\mathrm{Ga}}\right)=\left(-0.5,-3.0\right)$ eV, giving the results in Fig.~\ref{fig:form_energies} (d), for which $E_F^{SC}=0$ eV is found from charge neutrality. Under even more Ga-poor conditions, $E_F^{SC}$ remains pinned at 0 eV; the corresponding net hole concentration at 300 K is $p-n=7.8\cdot 10^{19}$ cm$^{-3}$.\\
\begin{figure}[h!!!t]
\centering
\includegraphics[height=6 cm]{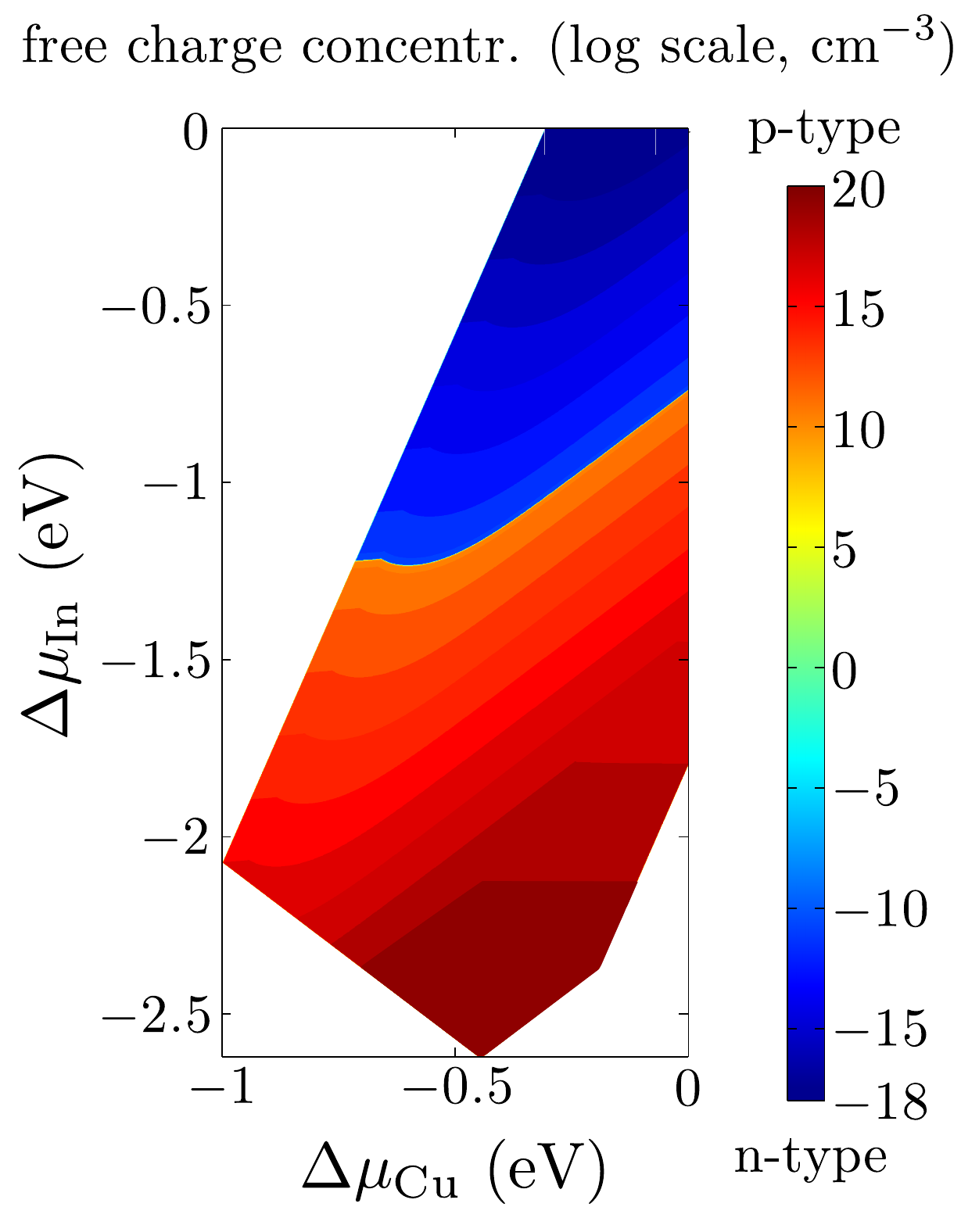}\llap{
  \parbox[b]{2.15in}{(a)\\\rule{0ex}{2.3in}
  }}
  
\vspace{0.3 cm}
\hspace{0.2cm}
\includegraphics[height=6 cm]{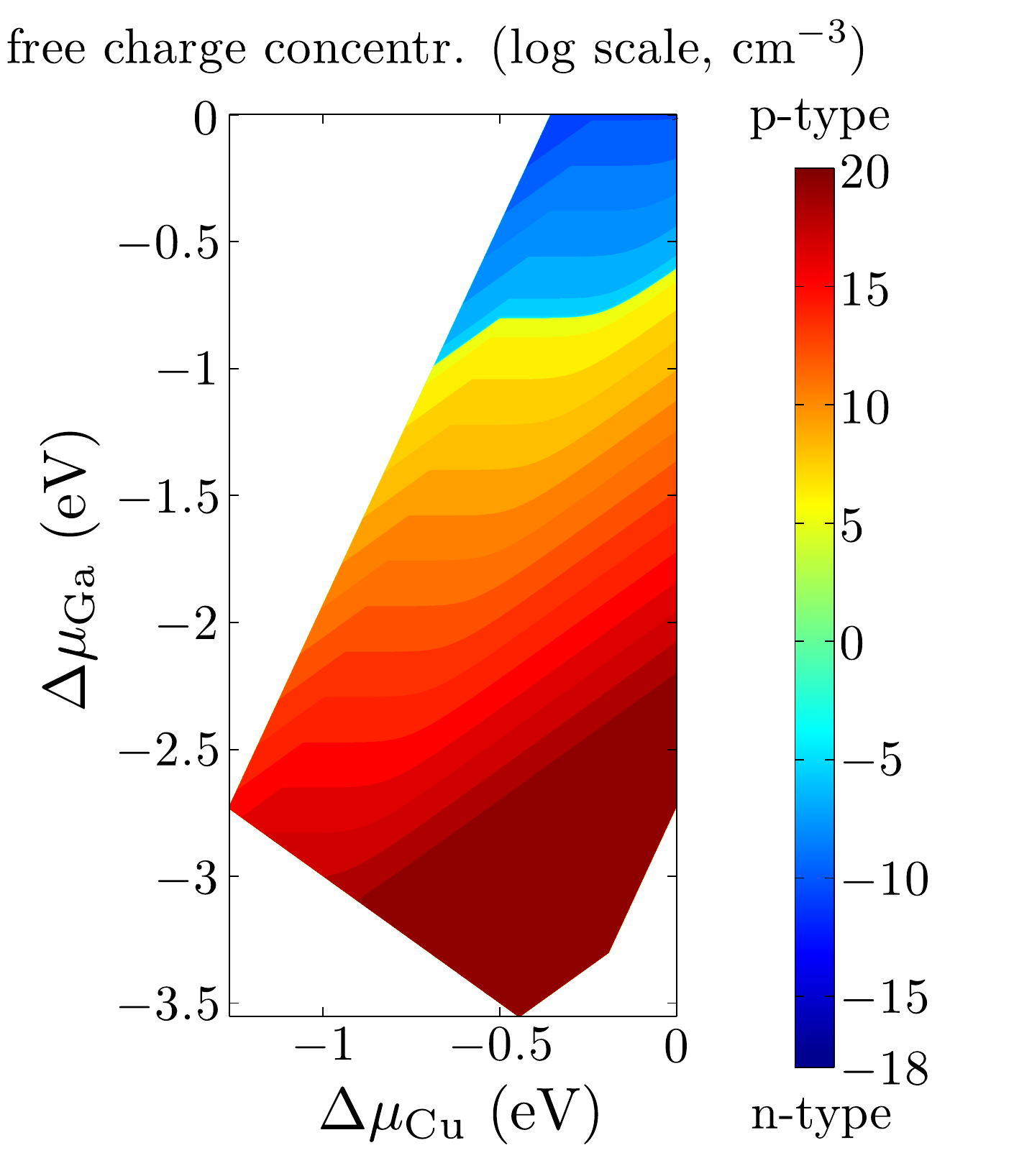}\llap{
  \parbox[b]{2.3in}{(b)\\\rule{0ex}{2.3in}
  }}
\caption{Contour plots of the net free charge carrier concentration $p-n$ (units of cm$^{-3}$ and on a log scale) in (a) CIS and in (b) CGS at a temperature of 300 K, as a function of $\Delta \mu_{\mathrm{Cu}}$ and $\Delta \mu_{\mathrm{In/Ga}}$ (for which CIS and CGS are stable). The positive values represent a net concentration of holes (p-type) and the negative values a net concentration of electrons (n-type).} 
\label{fig:contour}
\end{figure}
\indent From the study of the formation energies and resulting $E_F^{SC}$, we can draw a few important conclusions. First, the donors In$_{\mathrm{Cu}}$ and Ga$_{\mathrm{Cu}}$ are prevalent. In particular In$_{\mathrm{Cu}}$ causes strong n-type conductivity in CIS in In-rich conditions. In addition to this, In$_{\mathrm{Cu}}$ and Ga$_{\mathrm{Cu}}$ compensate to a large extent the acceptors. This results in potential fluctuations, measured via photoluminescence in Ref.~\citenum{Bauknecht,Siebentritt2004} in In-rich samples. The fluctuating potentials are reduced in In-poor samples, in agreement with the increasing formation energy of In$_{\mathrm{Cu}}$ and Ga$_{\mathrm{Cu}}$. Moreover, our results support the conclusion that the most shallow acceptor measured via photoluminescence in Ref.~\citenum{Siebentritt13} corresponds to V$_{\mathrm{Cu}}$, the second acceptor to Cu$_{\mathrm{In/Ga}}$ and the least shallow acceptor to V$_{\mathrm{In/Ga}}$. Namely, under increasingly Cu-poor conditions at fixed $\Delta \mu_{\mathrm{In/Ga}}$, we find that the formation energy of V$_{\mathrm{Cu}}$ increases (also taking into account changes in $E_F^{SC}$), while the formation energy of Cu$_{\mathrm{In/Ga}}$ increases and that of V$_{\mathrm{In/Ga}}$ is mostly constant. This provides a complete explanation for the changes in the intensities of the photoluminescence peaks due to the growth conditions, which we have discussed in the Introduction. In this way we can identify the different defect peaks without knowing the hydrogen-like defect levels. Calculating these would require supercells of sizes comparable to the Bohr radius, for CIS estimated to be $\sim$ 75~\AA, from photoluminescence measurements \cite{Yakushev}. Supercells of this size can currently hardly be used in DFT, especially in combination with the computationally demanding hybrid functionals. Another important consequence of our first-principles calculations is that under In-poor conditions, the antisite defects Cu$_{\mathrm{In}}$ and Cu$_{\mathrm{Ga}}$ have the lowest formation energy (form most easily) among the acceptors, rather than V$_{\mathrm{Cu}}$. This result explains why the concentration of holes increases with the stoichiometry $\left[\mathrm{Cu}\right]/\left[\mathrm{In}\right]$ and $\left[\mathrm{Cu}\right]/\left[\mathrm{Ga}\right]$ in CIS and CGS respectively \cite{Siebentritt13,Siebentritt01}, as we have already mentioned in the Introduction. This could not be explained if V$_{\mathrm{Cu}}$ were the principal acceptor defect in this case (as suggested in Ref.~\citenum{Wei2013}). In more Cu-poor and less In/Ga-poor conditions - near the edge adjacent to the CuIn$_5$Se$_8$ and CuGa$_5$Se$_8$ regions of the stability triangles - V$_{\mathrm{Cu}}$ is the dominant acceptor defect. Cu-rich maximum growth conditions are often avoided for the synthesis of CIGS for photovoltaic absorber layers, since the high $p$ concentration results in a narrow depletion region near the p-n junction \cite{Siebentritt13}. This is detrimental to the device performance because it enhances recombination near the interface. On the other hand, the transport and life time properties of Cu-rich absorbers are observed to be superior.\\
\indent The analysis of the influence of the chemical potentials on the conductivity type and related concentration can be extended to the whole chemical potential range, where CIS and CGS are stable - see Fig.~\ref{fig:triangles}. To this end, we have determined the self-consistent Fermi level yielding $p-n$ for each couple $\left(\Delta \mu_{\mathrm{Cu}},\Delta \mu_{\mathrm{In/Ga}}\right)$, in steps of 1 meV. We plot $\mathrm{sgn}(p-n)\cdot \mathrm{log}(\vert p-n \vert)$ in Fig.~\ref{fig:contour}, giving negative values in case of n-type conductivity and positive values for p-type. The contour plot for CIS shows that both n-type and p-type conductivity can easily be obtained, under respectively In-rich and In-poor conditions respectively. The maximum charge carrier concentration amount to $\sim 10^{18}$ cm$^{-3}$ for n-type and $\sim 10^{20}$ cm$^{-3}$ for p-type. The two types of conductivity have been realized in experiments, \textit{i.a.}~in Ref.~\citenum{Kristensen}. For CGS, Fig.~\ref{fig:contour} shows that it is much harder to establish n-type conductivity, prediciting concentrations limited to $\sim 10^{11}$ cm$^{-3}$, due to its wider band gap. Furthermore, we find lower charge concentrations around the n- to p-type transition, as the concentrations according to Eq.~\ref{eq:chcarrconcentr} diminish with $E_F$ deeper within the band gap. The very low concentration of free electrons (limited to $ 10^{11}$ cm$^{-3}$) we find theoretically demonstrates why - to our best knowledge - n-type undoped CGS has not been observed to date in experiments. It should be noted that n-type CGS has been realized, for instance by doping with Zn and Ge \cite{Schoen2000bis}. Regarding p-type conductivity, the highest net hole concentrations ($10^{18}$ to $10^{20}$ cm$^{-3}$) occur in a wider chemical potential range in CGS in comparison with CIS. This explains why in experiments the measured hole concentration increases with the Ga-to-In ratio \cite{Schroeder,Schoen2000}. Furthermore, the Fermi level is pinned at the VBM in p-type CGS under Ga-poor conditions, while it lies further in the band gap in CIS (\textit{cfr.}~Fig.~\ref{fig:form_energies}). Using Eq.~\ref{eq:chcarrconcentr}, we can also calculate how $p$ and $n$ evolve with temperature, thereby also taking into account the change in $E_F^{SC}$. As can be observed in Fig.~\ref{fig:freeze_out}, we predict a charge carrier freeze out between 500 and 50 K of less than one order of magnitude for CGS. In CIS, we find a larger freeze out, also depending more strongly on the chemical growth conditions. The predicted difference in magnitude of the freeze out between CIS and CGS corroborates experimental studies, such as Ref.~\citenum{Schoen2000}. The important effect of the chemical growth conditions on the freeze out in CIS can be linked to the variety in freeze out behavior in different experimental studies \cite{Schoen2000,Schumann1978,Matsushita1992}.
\begin{figure}[t]
\centering
\includegraphics[height=6.5 cm]{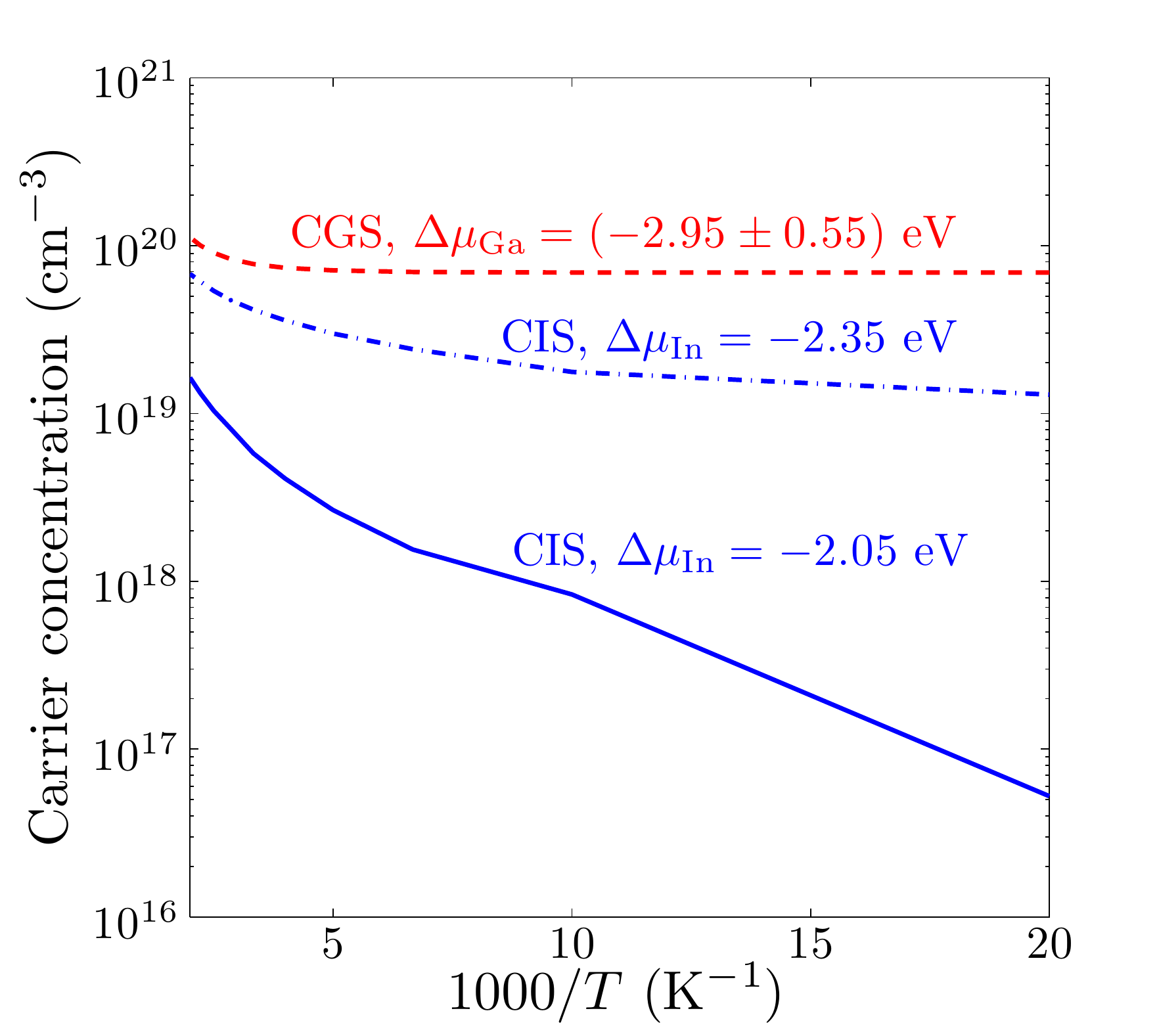}
\caption{The charge carrier concentration $p-n$ (cm$^{-3}$), on a logarithmic scale, as a function of 1000/$T$ with $T$ the temperature in K, thus representing the charge carrier freeze out between 500 K and 50 K. The chemical potential deviation for Cu is chosen $\Delta\mu_{\mathrm{Cu}}=-0.3$ eV in all cases, while we show the dependence on the chemical potential of In and Ga. The freeze out in CGS remains constant within the range $\Delta\mu_{\mathrm{Ga}}=(-2.95\pm0.55)$ eV due to the pinning of $E_F^{SC}$ at the VBM, whereas for CIS we find a strong difference between \textit{e.g.} $\Delta\mu_{\mathrm{In}}=-2.35$ eV and $\Delta\mu_{\mathrm{In}}=-2.05$ eV.}
\label{fig:freeze_out}
\end{figure}

\section{Conclusions}

We have obtained a good agreement with the experimental band gaps of CIS and CGS by using the hybrid HSE06 functional, with slightly enhanced intermixing of Hartree-Fock exchange interaction. From a band alignment of CIGS compounds using the branch-point energy concept, we conclude that the band gap mainly opens with increasing Ga-to-In ratio due to the rise of the conduction band minimum. This already indicates that the properties of p-type conductivity are similar in CIS and CGS, but n-type conductivity is much harder to establish in CGS. We have investigated the conductivity in undoped CIS and CGS by calculating the formation energy of native point defects in thermodynamic equilibrium. Our calculations show that In/Ga$_{\mathrm{Cu}}$ is a shallow donor, while V$_{\mathrm{Cu}}$, V$_{\mathrm{In}/\mathrm{Ga}}$ and Cu$_{\mathrm{In}/\mathrm{Ga}}$ act as shallow acceptors. Then, we have determined the Fermi level in the band gap - related to the exchange of electrons with charged defects - from charge neutrality, yielding the net free charge carrier concentration. The ionized defect concentrations were obtained by a Boltzmann distribution of the formation energies and the electron and hole concentrations were calculated from the density of states of CIS and CGS. This analysis reveals that the native donor In$_{\mathrm{Cu}}$ leads to strongly n-type conductivity in CIS in In-rich growth conditions. Under In-poor growth conditions the conductivity in CIS alters to p-type, while there is still compensation between donor and acceptor type defects (also found in experiment, \textit{i.a.}~Ref.~\citenum{Bauknecht,Siebentritt2004}). It diminishes under increasingly In-poor conditions as the formation energy of In$_{\mathrm{Cu}}$ goes up. In CGS, in contrast to CIS, we find a very low net concentration of electrons in n-type conditions (below $ 10^{11}$ cm$^{-3}$), owing to the Fermi level being pinned far away from the conduction band minimum by the native acceptors. This corroborates the absence of undoped, n-type CGS in experiments. On the other hand, CGS shows strong p-type conductivity (concentrations of $10^{18}$ to $10^{20}$ cm$^{-3}$) in a wider chemical potential range than CIS. Accordingly, a higher hole concentration is measured with increasing Ga-to-In ratio \cite{Schroeder}. Finally, our calculations lead to the conclusion that Cu$_{\mathrm{In}/\mathrm{Ga}}$ is the principal acceptor, with the lowest formation energy in CIS and CGS grown under In- and Ga-poor conditions. This explains why the hole concentration in experiment is found to be higher in samples with an In- and Ga-poor stoichiometry \cite{Siebentritt13,Siebentritt01}.

\section*{Acknowledgement}

We gratefully acknowledge financial support from the science fund FWO-Flanders through project G.0150.13. The first-principles calculations have been carried out on the HPC infrastructure of the University of Antwerp (CalcUA), a division of the Flemish Supercomputer Centre (VSC), supported financially by the Hercules foundation and the Flemish Government (EWI Department). We also like to thank Prof.~S.~Siebentritt of the University of Luxembourg for a presentation of her work on CIGS during a visit to our research group and for helpful discussions of our results.

\newcommand{\Ch}{Ch}

\footnotesize{
\bibliography{rsc} 
\bibliographystyle{rsc} 
}

\end{document}